\documentclass[preprint,5p,times,twocolumn,authoryear]{elsarticle}
\usepackage[utf8]{inputenc}
\usepackage[T1]{fontenc}
\usepackage{textcomp}
\usepackage{booktabs}

\usepackage{amssymb}
\usepackage{amsmath}

\usepackage{multirow}
\usepackage{algorithm}
\usepackage{algpseudocode}
\usepackage{lipsum}
\usepackage{chemfig}
\usepackage[hyphens]{url}
\setchemfig{atom sep=2.5em}
\newcommand{\Calpha}{\textcolor{red}{C}}
\newcommand{\SO}{\mathrm{SO}}
\usepackage{enumitem}
\usepackage{subcaption}

\usepackage[skip=2pt]{caption}

\journal{}

\begin{document}

\begin{frontmatter}

\title{Multiscale reconstruction of protein conformations from cryo-EM
images}

\author[label1]{David Y. W. Thong\corref{cor1}} 
\ead{dthong@kth.se}
\author[label1]{Ozan Öktem}
\author[label1]{Joakim Andén \corref{cor1}}
\ead{janden@kth.se}
\cortext[cor1]{Corresponding author}
\affiliation[label1]{ organization={Department of Mathematics, 
      KTH - Royal Institute of Technology},
  addressline={Lindstedtsvägen 25}, 
  postcode={100 44}, 
  city={Stockholm},
  country={Sweden}}

\begin{abstract}
We present a novel multiscale algorithm for directly recovering the atomic model structure of a protein from single-particle cryo-EM data. Our algorithm is able to estimate protein structures to state-of-the-art accuracy for high-noise and low-contrast data. It is also robust to misspecifications in the TEM image formation model. These desirable properties are primarily due to the use of an explicit representation of the protein backbone in terms of bonds, torsion angles and bond angles, which supplies rich prior information to the structure recovery process. We apply our method on three protein cryo-EM datasets, generated using an electron microscope digital twin, and show that using a multiscale approach yields an improvement of the root-mean-square deviation (RMSD) and template modelling (TM) scores with respect to the ground truth. Furthermore, there is evidence that larger-scale structures are being prioritised with the multiscale algorithm, which reduces the possibility of convergence to bad local minima.

\end{abstract}


\end{frontmatter}

\section{Introduction}

The 3D structure of biological macromolecules (such as proteins, nucleic acids, and complex macromolecular assemblies) is essential for understanding their function in sustaining cellular processes and diseases.
Cryogenic electron microscopy (cryo-EM) is a powerful method, capable of imaging large proteins in their natural state without the need for crystallisation, in which an aqueous sample containing many copies of the protein is flash-frozen \citep{cheng_primer_2015,milne_cryo-electron_2013} and then imaged using phase-contrast transmission electron microscopy (TEM). 
This produces hundreds of thousands of noisy 2D \emph{particle images}, each representing the projection of the (single) protein along an unknown direction. 

There are several major challenges associated with protein structure determination through cryo-EM. These include the fact that the obtained particle images have low-contrast, a very low signal-to-noise ratio, typically 0.1--0.01, and the process of determining atomic structure is sensitive to estimation of the \emph{contrast transfer function (CTF)} that models the TEM optics. 

Furthermore, traditional methods of signal/image processing in cryo-EM split the process into two steps: volume \emph{reconstruction}, in which a 3D volume of the protein is estimated from the 2D images, and \emph{model building}, in which an atomic structure is estimated from the volume. Note that the 3D volume from the first step is used in the second step, not the 2D particle images. As a result, these methods do not allow the integration of molecular structural information during reconstruction, and there is error propagation from the volume reconstruction step to the model building step.

The paper proposes a new \emph{multi-scale method for joint reconstruction and model building} that addresses these challenges. This algorithm performs simultaneous reconstruction and model building, which reduces error propagation, as opposed to most earlier methods, which split the process into two separate processes.
This joint approach also allows incorporation of structural constraints, which in two-step processes is not possible. Our approach expresses the protein backbone in terms of bonds and bond and torsion angles, analogous to internal coordinates. This reduces the possibility of physically implausible conformations being obtained and the need for tuning regularisation parameters.
 In addition, it makes our approach highly robust against misspecification of imaging parameters.

 In this method, we use anisotropic Gaussian basis functions for volumetric representation, which balances the expressibility and computational efficiency of the representation. 
Our approach is multiscale, which allows escape from local minima and prioritises large-scale structures in the early phases of the process (and thus reduces noise-related errors). Using the properties of the Gaussian basis (closed form under convolution) our multi-scale approach uses a coarse-to-fine optimisation approach. Analogous to pyramid techniques for image registration and scale space theories, our multiscale approach creates a hierarchical representation of the protein volume, creating lower resolution versions using convolution with a Gaussian kernel from higher resolution versions, so that an atomic model is fitted to larger features are fitted before smaller scale features. We validate our approach with realistic scenarios generated using a electron microscope digital twin showing the method's ability to estimate accurate structures.

The following sections will give a brief overview of the existing literature within cryo-EM for reconstruction and model building and how our method fits within this. Then we will go over the methods and their mathematical framework. Finally, we will validate the method on several scenarios generated with realistic settings with a electron microscope digital twin.

\section{Current state-of-the-art}

This section gives a brief overview of previous work, including both two-stage and joint methods, their existing limitations, and how our method addresses these limitations.

\subsection{Existing volume reconstruction methods}
Volume reconstruction (hereafter just \emph{reconstruction}) refers to mathematical and computational methods for recovering a 3D map, or \emph{volume}, depicting the protein, from the 2D particle images. Many software packages implement various approaches for volume reconstruction and notable ones include RELION \citep{scheres_relion_2012}, CryoSPARC \citep{punjani_cryosparc_2017}, CryoDRGN~1 and 2 \citep{zhong_cryodrgn_2021,zhong_cryodrgn2_2021}, E2GMM \citep{chen_deep_2021}, 3DFlex \citep{punjani_3dflex_2023}, CryoGAN \citep{gupta_cryogan_2021}, DynaMight \citep{schwab_dynamight_2024}, CryoAI \citep{levy_cryoai_2022}, CryoFold \citep{zhong_exploring_2021}, Multi-CryoGAN \citep{gupta_multi-cryogan_2020}, and CryoPoseNet \citep{nashed_cryoposenet_2021}. These implementations deal differently in handling the unknown poses and they may also have different representations of the 3D map.

There is no representation of the molecular structure in volume reconstruction, which makes it difficult, or impossible, to impose physical or structural constraints. As a consequence, one can only set volume priors, like smoothing/regularity, whereas priors at atomistic level cannot be imposed. As a result, regions that involve fine details, or have low resolution, like flexible regions, appear as vague and ambiguous densities. In addition, noise can be mis-interpreted as part of the density. This occurs in all approaches, but with no priors at the atomic level, it is more likely the physically impossible densities may be obtained from the reconstruction process.

\subsection{Existing model-building methods}
Early approaches, like COOT \citep{emsley_features_2010}, entailed fitting atomic models to cryo-EM density maps by hand, or used semi-automated approaches guided by strong human intervention as in PHENIX's Autobuild \citep{adams_phenix_2010} and Buccaneer \citep{cowtan_buccaneer_2006}.
These methods originated from X-ray crystallography methods, and often required large amounts of biochemistry and model-building experience. 

Recent years have seen the appearance of several model-building methods, both traditional approaches and others based on deep learning. An example of a traditional approach is Phenix \citep{liebschner_macromolecular_2019, afonine_real-space_2018, terwilliger_iterative_2008} and examples of deep-learning based approaches are ModelAngelo \citep{jamali_graph_2023, jamali_automated_2024}, CR-I-TASSER \citep{zhang_cr-i-tasser_2022} and DeepMainmast \citep{terashi_deepmainmast_2024}. These either impose structural constraints through regularisation or explicit constraints, or learn structural constraints from large datasets.

The above cited methods, although sophisticated, do not use the raw image data obtained from the cryo-EM process. They are a post-processing step that is applied to a 3D map obtained from a preceding volume reconstruction step. Hence, any errors from this previous step will propagate into the model-building.

\subsection{Joint volume reconstruction and model-building}
The previous mentioned sequential two-step methods has a separation between model-building and volume reconstruction that prevents information from passing between the two tasks. This is not the case with joint volume reconstruction and model-building where these two tasks are done simultaneously.

There are some approaches for joint reconstruction and model-building, e.g., \citet{li_cryostar_2024}, \citet{zhong_exploring_2021}, \citet{rosenbaum_inferring_2021}, \citet{chen_integrating_2023}, and \citet{ducrocq_cryosphere_2024}. The joint approach has the potential to reduce, or even prevent, the propagation of error from the volume reconstruction step into the model-building step. In addition, since these methods allow explicit modelling of the atomic structure, they also open up for usage of more sophisticated priors.

Existing approaches model atoms in the protein as a point cloud, with the potential from each atom represented by Gaussian basis functions. Enforcing structural constraints serves as a regularisation. These methods have been shown to be highly effective on a variety of synthetic or experimental data sets, e.g., both \citet{chen_integrating_2023} and \citet{li_cryostar_2024} have shown strong performance on experimental data. 
Both of these approaches can be thought of as adopting a ``coarse-to-fine'' approach\footnote{A coarse-to-fine approach is one in which a problem is expressed as a hierarchy of sub-problems, going from coarse-scale problems (which are restricted versions of the problem, for example, a version of the problem with lower-resolution, or fewer parameters), to fine-scale/full versions of the problems (which are versions of the problem where these restrictions are progressively removed). Information is transferred up the hierarchy, from the coarser scale to the finer scales up to the full version of the problem.}.

\citet{chen_integrating_2023} uses a Gaussian mixture models (GMM) with two Gaussians, a high-resolution GMM and a low-resolution GMM, where the former consists of tens of thousands of Gaussians, the latter of hundreds. Conformations of the protein volume are here represented as deformations of the high-resolution GMM fitted to the template structure. These deformations of the high-resolution GMM are parametrised by a low resolution GMM, the position of the centres of each element of the low resolution GMM acting as control points on the deformation of high-resolution GMM through a transition matrix. There are two phases: first, the transition matrix is held fixed, with the centres of the low resolution GMM being related to the high resolution GMM centres through a function of distance, in the second phase, the transition matrix is learnt.

\citet{li_cryostar_2024} does not have multiscale elements in the usual sense, but it nevertheless employs a coarse-to-fine approach in the form of adaptive regularisation of structural constraints. A regulariser enforces an elastic network between all atoms within a given distance. During each training iteration, if the intra-batch variance of a pair of interatomic distances is high, the elastic network connection between the pair of atoms is removed thereby allowing a greater range of motion for the atoms. Hence, initially the structural regulariser will enforce rigidity, with the constraint being relaxed in areas where more flexibility is needed.

These methods have several features in common, the most prominent one being that structural constraints and molecular structure is not explicitly modelled, making it more likely that the optimisation process may result in conformations that are not biophysically possible. In addition, results of these methods are sensitive to choice of these hyperparameters, e.g., one needs to set the strength of the regularisation.

Finally, most methods rely on isotropic Gaussians \citep{li_cryostar_2024, ducrocq_cryosphere_2024, rosenbaum_inferring_2021, zhong_exploring_2021}, which have limited flexibility in efficiently expressing the potential generated by the molecule. Furthermore, all methods require modelling the TEM optics by specifying the CTF, and the quality of results depends on the how well one estimates the CTF parameters.

In conclusion, approaches for finding the atomic structure for a given protein from cryo-EM image data has been, in the majority of approaches in the past, a sequential two-stage process: volume reconstruction and model building. Recently, there have been approaches at performing joint reconstruction and model building. However, there has been a paucity of approaches using a multiscale approach, and explicitly integrating prior structural information (instead of through regularisers in optimisation). Our contribution is a multiscale method for joint reconstruction and model building that integrates prior information about the molecular structure and chemical composition. The volume is here expressed through an anisotropic Gaussian dictionary that is linked to the atomic positions. This allows joint reconstruction and model building where atomic structure of the protein also explicitly encodes not only atomic positions, but also covalent bonds and bond angles, thus allowing the optimisation process to move within the manifold of feasible conformations.

\section{Mathematical model}
We here describe the mathematical framework for representing conformations of protein structures.
Specifically, we propose a method for parametrising the conformation of the protein backbone chain in terms of torsion and bond angles. Using this parametrisation, the optimisation is performed on a manifold that represents conformations where covalent bonds are preserved. We then go on to describe the mathematical framework for linking a conformation to a volume, which then can be used as input by an TEM image formation model (forward operator). Finally, we will describe the multi-scale algorithms that are used for joint reconstruction and model building.

\subsection{Problem statement}
Joint reconstruction and model-building can be seen formally as optimising the data fidelity between the observed 2D TEM images $\{\mathbf{y}_i\}$ and synthetic 2D TEM images that are generated from a point cloud representing the atoms in the protein. The conformations generating these 2D TEM images are assumed to be deformations of a template conformation. Stated formally, we seek to find the conformation parameters that minimises
\[ 
\Delta\psi \mapsto \sum_i \left\Vert \mathcal{F}_{\omega_i}\circ\mathcal{G}\circ \mathcal{W}_{\Delta\psi}(B_0)-\mathbf{y}_i \right\Vert^2_2,
\]
where $B_0$ is a known set of coordinates obtained from the template structure, $\Delta\psi$ is a variable that describes the protein conformation (in relation to the template structure), $\mathcal{W}_{\Delta\psi}$ is an operator that transforms the atomic coordinates according to $\Delta\psi$ (which corresponds to a change in conformation), $\mathcal{G}$ is an operator that generates a volume from an atomic structure, and $\mathcal{F}_{\omega_i}$ converts a volumes with pose angle $\omega_i$ into a 2D TEM image. Further details of how we define these in our method will be given later. This formulation intends to capture the essential relationship between template structure, conformation parameters, and observed data while deferring implementation details to subsequent sections.

\subsection{Motivation}
The simplest method for performing the joint reconstruction and model-building, given a known \emph{template} structure as a starting point, would be to treat the atoms in the protein molecule as a point cloud, each atom producing a Gaussian 2D image when imaged using the electron microscope. This is compared, during optimisation, with the images obtained from the microscope. This direct approach has several challenges.

The first is that is difficult to optimise for protein structures which respect e.g. covalent bonding, when optimising the atomic positions. The second is that Gaussian basis functions concentrate over a small area, and thus if parts of the template structure do not overlap with the true structure, this produces problems with near-zero gradient and results in difficulty converging to the true structure. The third is that proteins have tens or hundreds of thousands of atoms, and if the 2D projection of such a number of 3D Gaussian basis elements were taken, there would be a large computational burden (both on memory and generating projections during each iteration of the optimisation process).

These challenges motivate our method, of which there are four major components. First of all, we explicitly model atomic positions in terms of bonds and bond angles, to constrain optimisation to conformations that respect covalent bonding. Secondly, our method utilises a multiscale representation of the 3D volume using anisotropic Gaussian dictionaries, to mitigate convergence problems. Thirdly, our method incorporates an algorithm to merge per-atom Gaussian basis functions into per-residue anisotropic basis functions. Finally our method uses an explicit relationship between the atomic position and multiscale volumetric representation. Our method is designed with these four components to meet the challenges described above.

\subsection{Mathematical framework}

In this section, we will describe the mathematical framework that is used in our method to describe protein chains, their relationship to the to the volume of the protein molecule, and the relationship to the particle images.

\subsubsection{Protein chains}

A protein comprises one or more protein chains, each of which consists of amino acids connected via peptide bonds. All amino acids consist of an amine group, a carboxyl group and a side chain. Many such amino acids, linked by peptide bonds, create what is known as the backbone chain.

The protein chain consists of many side chains attached to the backbone. Each side chain is attached to a carbon atom in the backbone chain, known as the $C_\alpha$ (carbon alpha) atom. The $C_\alpha$ atom is bonded to the nitrogen of the amine group (from the previous residue’s peptide bond), the side chain, and the carboxyl carbon (which forms a peptide bond with the next residue’s nitrogen).

\begin{figure}[tb]
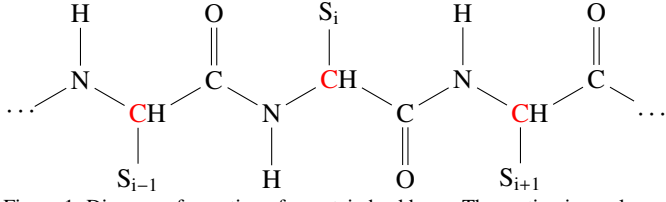

    \centering
    \chemfig{
\cdots%
-[:30]%
N(-[2]H)%
-[:-30]%
\Calpha H(-[6]S_{i-1})%
-[:30]%
C(=[:90]O)%
-[:-30]%
N(-[-2]H)%
-[:30]%
\Calpha H(-[-6]S_i)%
-[:-30]%
C(=[:-90]O)%
-[:30]%
N(-[2]H)%
-[:-30]%
\Calpha H(-[6]S_{i+1})%
-[:30]%
C(=[:90]O)%
-[:-30]%
\cdots
}

    \caption{Diagram of a section of a protein backbone. The section is a polypeptide made of three amino acids polymerised together. Each amino acid in the chain differs by its side chain, denoted $\mathrm{S}_{\mathrm{i-1}},\mathrm{S}_{\mathrm{i}},\mathrm{S}_{\mathrm{i+1}}$. The $C_\alpha$ atoms in the chain are highlighted in red. }
    \label{fig:backbone-struct}
\end{figure}

In this way, the backbone of a protein chain of $n$ residues can be thought of as comprising triplets of atoms.
More precisely: by backbone chain, we refer to the heavy atoms: the nitrogen $N$, which is part of the amine group, carbon $C_{\alpha}$ to which a side chain is attached (labelled as $S_i$ in chemical diagrams), and a carbon atom $C$ in the carboxyl group (see Figure~\ref{fig:backbone-struct}).

The spatial location of an amino acid in the backbone can be represented by a triplet of atomic coordinates representing the position of the nitrogen ($N$), $C_\alpha$, and the carboxyl carbon ($C$) atoms, respectively. 
Hence, the conformation of a single protein backbone chain consisting of $n$ amino acids can be denoted by the array of $n$ triplets
\begin{equation}\label{eq:SeqOfTriplets}
B = \begin{bmatrix}
\bigl(
\mathbf{r}_{0}, 
\mathbf{r}_{1}, 
\mathbf{r}_{2}
\bigr)
&
\ldots
&
\bigl(
\mathbf{r}_{3(n-1)}, \mathbf{r}_{3(n-1)+1}, \mathbf{r}_{3(n-1)+2}
\bigr)
\end{bmatrix}
\in \mathbb{R}^{(3 \times 3)n}
\end{equation} 
where $(\mathbf{r}_{3i},\mathbf{r}_{3i+1},\mathbf{r}_{3i+2}) \in \mathbb{R}^{3\times 3}$ encodes the positions of the nitrogen ($N$), $C_\alpha$, and the carboxyl carbon ($C$) atoms in the $i$th amino acid for $i=1,\ldots, n-1$.

This is the most basic representation of the protein that we will use in the current work.
It is comprehensive in that it describes the conformation of the backbone, but it is not useful for optimisation since moving one of these atomic coordinates independently from the others would result in breaking the covalent bonds of the chain.
We will instead consider deformations (conformational changes) that preserve bond lengths.

\subsubsection{Manifold of protein backbone conformations}\label{sec:manifprot}
Although protein structures can change, there are several invariants, and several other elements which can be assumed to be approximately invariant. Changes in structure are primarily determined by the conformation of the backbone chain, and covalent bond lengths remain approximately constant. 
Searching for conformations of a given protein can therefore be restricted to searching on the manifold of ``feasible'' conformations. 
Each point on this manifold represents a possible backbone conformation and it is given by $n$ triplets as in \eqref{eq:SeqOfTriplets} subject to constraints of fixed bond lengths and backbone chain ordering. 
In other words, this manifold of valid protein backbone chain configurations is a sub-manifold to the space of $n$ triplets that has much lower dimension, so the computational cost in searching for a conformation is much reduced by restricting the search to this valid protein backbone chain configurations. 
This also avoids some backbone conformations that are impossible or unlikely. 
We impose a mathematical structure on these triplets of coordinates by leveraging the fact that a protein backbone chain is a peptide chain linked by covalent bonding, and we assume the covalent bonds do not break.
These bonds can be assumed to be of constant length (as is commonly done in protein molecular dynamics).

To formalise the above, we define the \emph{manifold of backbone conformations} for a protein with $n$ residues as 
\[ \mathcal{M}_n = \mathbb{R}^3 \times \mathrm{SO}(3) \times [0,\pi]^{3n - 2} \times (-\pi,\pi]^{3n - 2}. \]
Hence, a point on $\mathcal{M}_n$ is a tuple $\bigl( \mathbf{r}_0, F_0, \{ \kappa_i \}_{i=1}^{3n-2}, \{ \tau_i \}_{i=1}^{3n-2} \bigr)$ that encodes the location of the first atom $\mathbf{r}_0$, the orientation of the first bond $F_0$, plus the bond and torsion angles $\{\kappa_i\}_{i=1}^{3n-2}$ and $\{\tau_i\}_{i=1}^{3n-2}$, respectively, of the remaining bonds. See \ref{app:manif} for more details.

\begin{figure}[tb]
    \centering
    \includegraphics[width=\linewidth]{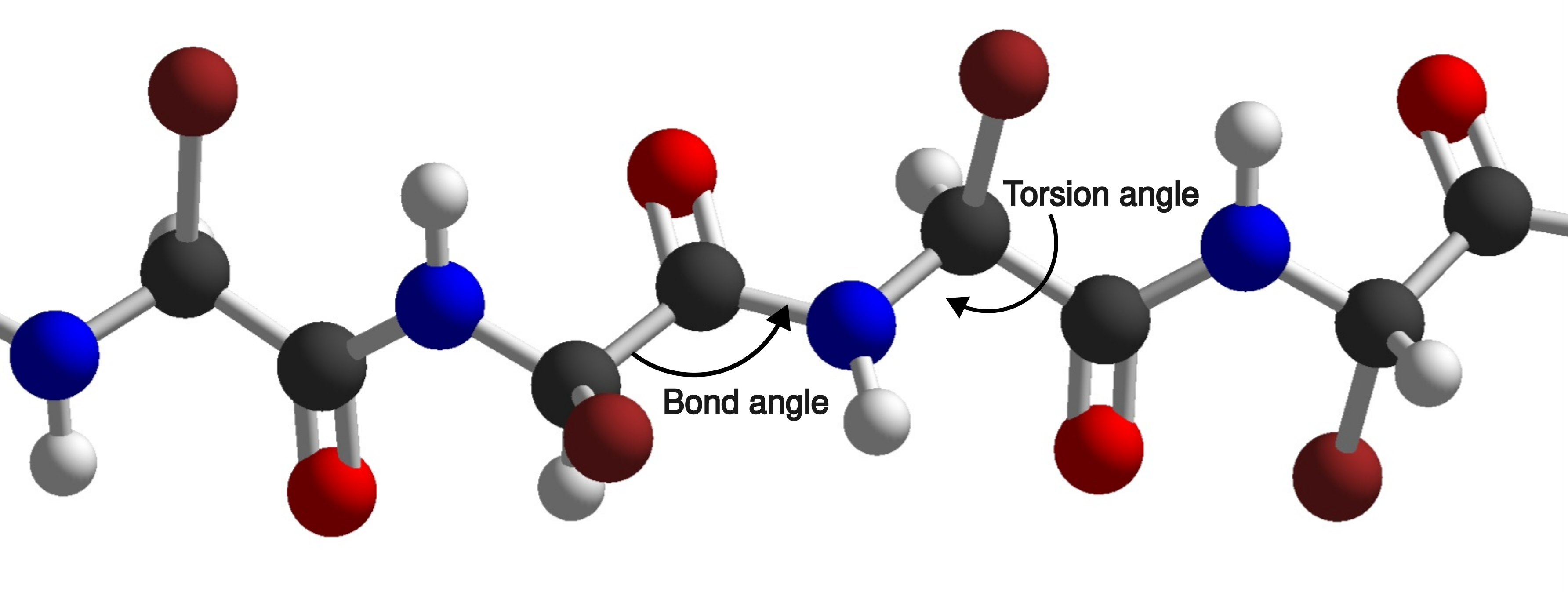}
    \caption{$\mathcal{W}_{\Delta\psi}$ is parametrised by $\Delta\psi$ which includes the change in initial Frenet frame $\Delta F_0$, bond angles $\{\Delta \kappa_i \}_{i=1}^{3n-2}$ and torsion angles $\{\Delta \tau_i \}_{i=1}^{3n-2}$ of the deformed backbone}
    \label{fig:rot-torsions}
\end{figure}

Let $\psi_0=\bigl( \mathbf{r}_0^{\;0}, F_0^{\;0},\{ \kappa_i^{\,0} \}_{i=1}^{3n-2},\{ \tau_i^{\,0} \}_{i=1}^{3n-2} \bigr)\in \mathcal{M}$ be the template conformation.
The mapping $\mathcal{W}_{\Delta\psi} \colon \mathcal{M} \to \mathcal{M}$ (deformation operator) that models the change in conformation can now be parametrised by $\Delta\psi \in \mathcal{M}$. 
To see this, a conformation $\psi = (\mathbf{r}_0, F_0, \{\kappa_i\}_{i=1}^{3n-2}, \{\tau_i\}_{i=1}^{3n-2}) \in \mathcal{M}$ is deformed into $\psi'=\mathcal{W}_{\Delta\psi}(\psi)$ with $\Delta\psi = \bigl( \Delta\mathbf{r}_0,\,\Delta F_0,\,\{ \Delta\kappa_i \}_{i=1}^{3n-2},\,\{ \Delta\tau_i \}_{i=1}^{3n-2} \bigr) \in \mathcal{M}$ where $\psi' = \bigl(\mathbf{r_0} + \Delta\mathbf{r}_0, F_0\Delta F_0, \{\kappa_i + \Delta \kappa_i\}_{i=1}^{3n-2}, \{\tau_i + \Delta \tau_i\}_{i=1}^{3n-2}\bigr)$ (see Figure~\ref{fig:rot-torsions}).

\subsubsection{Continuous volumetric representation} \label{sec:local-frame}
We define $\mathcal{G} \colon \mathbb{R}^{(3\times 3)n} \to L^2(\mathbb{R}^3)$ to be the operator which takes the backbone coordinates $B$ of a protein chain and yields a real-valued function representing the volumetric representation of the protein chain. For the purpose of simplicity, we consider $\mathcal{G}$ to be defined by a linear combination of dictionary elements  $g_0,g_1,\ldots,g_{n-1} \colon \mathbb{R}^{3 \times 3} \times \mathbb{R}^{3}  \to \mathbb{R}$, i.e. for $B \in \mathbb{R}^{(3 \times 3)n}$ as in \eqref{eq:SeqOfTriplets} we have
\begin{equation}
    \mathcal{G}(B)(\mathbf{x})=\sum_{i=0}^{n-1} w_i \, g_i(\mathbf{x}, B)
\quad\text{for $\mathbf{x} \in \mathbb{R}^{3}$.}
\label{eq:gaussrepsum}
\end{equation}
where each Gaussian $g_i$ depends on three consecutive backbone atoms of residue $i$ of backbone $B$, and has weight $w_i$. The three backbone atoms are the $N,C_\alpha$, and $C$ atoms of each residue. Each $g_i$ thus represents a residue in the protein chain. As the side chain often comprises most of the mass of the residue, compared to the three backbone atoms, this Gaussian will not often be centred on the backbone chain.

Since the Gaussian representing each residue is not centred on the backbone chain, but rather, at a distance from it, we need a model of how the Gaussian should be positioned oriented relative to the protein backbone, as the Gaussian is anisotropic and may be centred at some distance from the backbone. The formal mathematical formulation is given in \ref{app:relgauss}.

\subsubsection{Anisotropic Gaussian dictionary \label{sec:gaussrep}}

We aim to perform optimisation over the space of all reasonable protein structures using explicit structural constraints. Since the conformation of a protein is primarily determined by its backbone, we have developed the mathematical framework, assuming constant bond lengths, that expresses a backbone in terms of a starting point, an initial Frenet frame and sequences of torsion and bond angles. These can be converted to atomic coordinates and vice-versa. 

As we optimise, we need to compare (through the use of the forward operator) the shape of the protein backbone against observed images. To do so, we express the electron density as a Gaussian dictionary, as in Eq. \ref{eq:gaussrepsum}, where each Gaussian dictionary element approximates the electron density of each residue, and create a mathematical framework which relates the current position of the Gaussian dictionary elements with the the chain's conformation. 

We need to determine the parameters of each Gaussian dictionary elements. These could be added as part of the optimisation process, but this would increase the number of parameters. Instead, we calculate them based on the physical properties of the atoms and residues. See \ref{app:gaussfus} for details of the mathematical formulation.

\section{Proposed method}

In this section, we will explain how the theory of the previous sections fits together, showing how joint reconstruction and model building is performed from start to finish. Suppose that we want to perform reconstruction and model building on cryo-EM data from a given molecule, and we have particle picked, cropped and centred images of the protein molecule, known poses, and we have a PDB file containing one known reference conformation of the molecule. The process consists of multiple algorithms, performed sequentially. First, the template is globally aligned to the image data using Algorithm~\ref{alg:glob} (a global alignment algorithm). This global alignment algorithm aligns the 3D template backbone to the 2D images, without any change of internal conformation, which greatly reduces computation time. Then, multiscale method described in Algorithm \ref{alg:mscalealg} is applied, which is the core of our method. We will explain these stages in further detail in the following sub-sections.

\subsection{Multiscale algorithm}

When performing optimisation for this problem, the optimisation algorithm can converge to poor local minima. To mitigate this, we use a multiscale approach, where a coarsened version of the volumetric representation is fitted to the image data. This is then iteratively refined over several stages, until the optimisation algorithm is run on data at the finest scale.

\begin{figure*}[tb]
    \centering
    \includegraphics[width=0.9\linewidth]{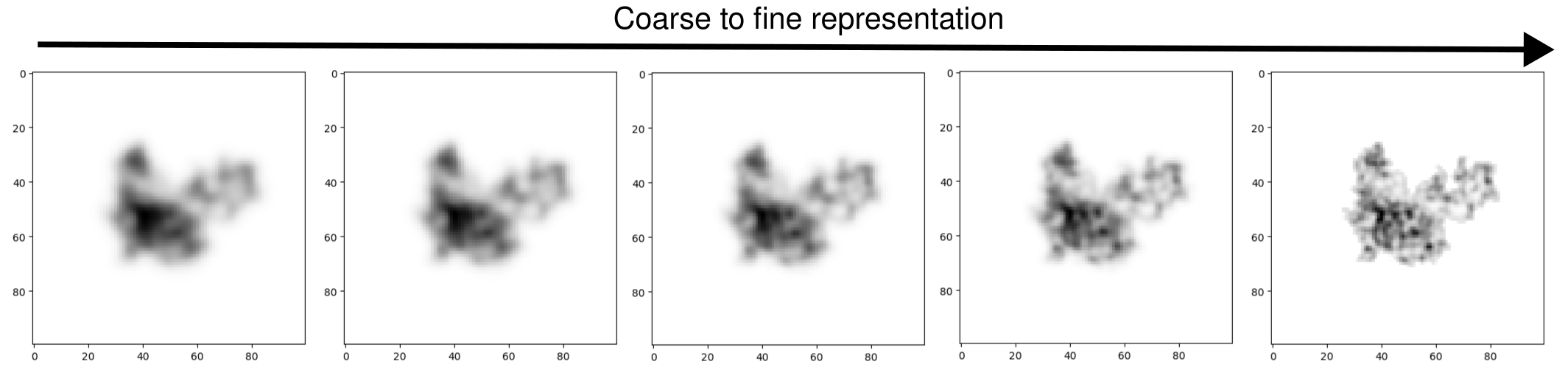}
    \caption{An iterative refinement approach, where optimisation is performed on a coarsened representation, and is repeatedly refined at increasing levels of detail.}
    \label{fig:iterative-refinement}
\end{figure*}

The volumetric representation is coarsened through convolution with a Gaussian kernel. As the convolution of a Gaussian with a Gaussian kernel is itself a Gaussian, this operation can be performed with minimal computational cost (see Figure~\ref{fig:iterative-refinement}).

The motivation for this method of coarsening comes from the use of the $L^2$ norm to compare the data with projections of the volumetric representation. As the majority of the mass of the Gaussian basis functions is concentrated over a small region, if two of the Gaussians do not overlap significantly, the $L^2$ norm has almost flat gradient at that point, and the algorithm will not be able to align the Gaussians. This method of coarsening lessens the effect by dispersing the Gaussians of the volumetric representation.

Another perspective of the effect of this coarsening is that the convolution of the volumetric representation with a Gaussian is that high spatial frequencies in the volume (and thus its projections) are attenuated, thus prioritising the fitting of large-scale elements over small-scale elements.

\subsection{Algorithmic details}

The input data consists of $N$ particle images, which are normalised to have mean zero and variance one. The PDB data is read, and by parsing the atomic coordinates and element data, we obtain $B_0$, as discussed earlier.
Using the element data in the PDB file and $B_0$, the process in Section~\ref{sec:gaussrep} is then used to obtain the volumetric function $\mathcal{G}(B)$, which is a sum of Gaussians tied to the protein's backbone conformation.

The next step is to define the manifold of protein conformations from the backbone coordinates of the template conformation, $B_0$. We define this as in Section \ref{sec:manifprot}, obtaining $\mathcal{M}_n$ as the manifold of protein conformations. Each $\psi = (\mathbf{r}, F_0, \{\kappa_i\}, \{\tau_i\})$  in this manifold represents a different backbone conformation, where the bond lengths remain the same as $B_0$. From $B_0$, we note the value  $\psi_0 = (\mathbf{r}_0, F_{0}^0, \{\kappa^0_i\}, \{\tau^0_i\})$, which will be used initialise the optimisation algorithms. We denote the change from this template conformation by $\Delta\psi = (\Delta\mathbf{r}, \Delta F_0, \{\Delta\kappa_i\}, \{\Delta\tau_i\})$.

\begin{algorithm}[bt]
\caption{Global alignment algorithm \label{alg:glob}}
\begin{algorithmic}
    \State \textbf{Input:} Template backbone coordinates $B_0$ parametrised by $\psi_0 =  (\mathbf{r}_0, F_{0}^0, \{\kappa^0_i\}, \{\tau^0_i\})$, target images $\{\mathbf{y}_i\}$, viewing directions $\{\boldsymbol{\omega}_i\}$, optimisation iterations $K$, number of residues $n$
    \State $\mathrm{BestLoss} \gets \sum_i \left\Vert \mathcal{F}_{\omega_i}\circ\mathcal{G}\circ\mathcal{W}_{\Delta\psi}(B_0)-y_i\right\Vert^2_2$
    \State $\Delta\psi_{\mathrm{best}} \gets \Delta\psi_0$
    \For{$k=1\ldots K$}
      \State $\Delta F_{0} \;\gets\; \text{Unif}(\SO(3))$ 
        \State \textbf{Stage A (optimise translation only):} \\ 

        \[
            f_\mathrm{trans}(\mathbf r) :=
            \sum_i \bigl\|\,
                \mathcal F_{\omega_i}\!\circ\!\mathcal G\!\circ\!
                \mathcal W_{(\Delta \mathbf r,\, \Delta F_{0} \cdot F_{0} ,\mathbf{0},\mathbf{0})}(B_0)
                -\mathbf y_i
                \,\bigr\|_2^{2}
        \]
      
        \State 
              $\displaystyle \Delta \mathbf r \gets 
                 \Call{StochasticGradientDescent}{f_\mathrm{trans},\,\Delta \mathbf r}$
                 
                 \Comment{$\Delta F_{0}$ is fixed}

        \State \textbf{Stage B (optimise translation and rotation):} \\ 
        \[
            f(\Delta\psi) :=
            \sum_i \bigl\|\,
                \mathcal F_{\omega_i}\!\circ\!\mathcal G\!\circ\!
                \mathcal W_{(\Delta \mathbf r, \Delta F_{0} \cdot F_{0},
                \mathbf{0},\mathbf{0})}(B_0)-\mathbf y_i
                \,\bigr\|_2^{2}
        \]
        \State 
              $\displaystyle \Delta\psi \gets 
                 \Call{StochasticGradientDescent}{f,\, (\Delta \mathbf r, \Delta F_{0})}$          
                 
                 \Comment{optimise $\Delta \mathbf r, \Delta F_{0}$}

    \If {$f(\Delta\psi)< \mathrm{BestLoss}$} \State $\mathrm{BestLoss} \gets f(\Delta\psi)$ \State $\Delta\psi_{\mathrm{best}} \gets \Delta\psi$
    \EndIf
    \EndFor
    \State $\Delta\psi\gets$ \Call{StochasticGradientDescent}{$f$,$\Delta\psi_{\mathrm{best}}$}
    \State \Return{Rotated and translated backbone conformation $\Delta\psi$}
\end{algorithmic}
\end{algorithm}

To improve speed of convergence and convergence to a good local minimum, a global alignment stage given by Algorithm~\ref{alg:glob} is first performed. The global alignment algorithm aligns a 3D template backbone to the set of 2D target images by optimising its parameters through a combination of random restarts and stochastic gradient descent~(SGD). The global alignment algorithm performs SGD to optimise the initial Frenet frame. The optimisation is performed on the $\Delta\mathbf{r}$ and $\Delta F_0$ components of $\Delta\psi$, which is equivalent to rigid body alignment. The optimisation is performed several times with random starting points. Throughout the process, the best parameters found are recorded. After all iterations, a final refinement is performed on the best parameters and the optimised backbone conformation that best matches the target images is returned.

The cryo-EM images and known pose angles are used with the forward operator (described in \ref{sec:volimg}) to define the loss function. The multiscale algorithm, Algorithm \ref{alg:mscalealg} is then used to optimise $\Delta\psi$.

\begin{algorithm}[bt]
\caption{Multiscale algorithm \label{alg:mscalealg}}
\begin{algorithmic}
    \State Initialisation: template molecular model $B_0$, target images $\mathbf{y}$, $s \gets 1$, maximum blur $\beta > 0$, $K$ number of multiscale levels
    \State $\Delta\psi_0 \gets$ \Call{GlobalAlignmentAlgorithm}{$B_0, \mathbf{y}$}
    \For{$s=1\ldots K$}

    \If{$k < K$}
        \State $\sigma \gets \beta - (s-1)\,\dfrac{\beta -0.75}{K - 2}$
    \Else
        \State $\sigma \gets 0$
    \EndIf
        \State $\mathcal{G}_{\sigma}:=\mathcal{K}_{\sigma}\star\mathcal{G}$ where $\mathcal{K}_{\sigma}$ is an isotropic Gaussian kernel of bandwidth $\sigma$.
        \State $f(\Delta\psi):=\sum_i \left\Vert \mathcal{F}_{\omega_i}\circ\mathcal{G}_{\sigma}\circ \mathcal{W}_{\Delta\psi}(B_0)-\mathbf{y}_i\right\Vert^2_2$
        \State $\Delta\psi_s\gets$ \Call{StochasticGradientDescent}{$f$,$\Delta\psi_{s-1}$}
    \EndFor
    \State \Return{Backbone $B = \mathcal{W}_{\Delta\psi_s}(B_0)$}
\end{algorithmic}
\end{algorithm}

The algorithm begins by taking the approximately aligned molecular structure from the global alignment algorithm. The volumetric representation is convolved with an isotropic Gaussian kernel, which ``spreads out'' or ``blurs'' it. This blurred volumetric representation is fitted to the images by optimising using SGD over the torsion bond angles, start coordinate and initial Frenet frame. 

The process is repeated at progressively finer scales: the blur $\sigma$ is set to a smaller value, so the Gaussians are less dispersed and finer details become visible. The algorithm is initialised with the results of the optimisation from the previous stage, transferring information from the previous coarser stage to the current finer stage. This is repeated at finer scales until there is no blur. This coarse-to-fine approach encourages the algorithm to align large scale structural features first (e.g. opening and closing of the molecule) and  refines by aligning smaller structural features later (e.g. movement due to vibration). This avoids poor local minima.

We have implemented this using the PyTorch library \citep{paszke_pytorch_2019}, which allows for automatic differentiation and evaluation on the GPU.

\section{Numerical experiments \label{sec:num-exp}}

\begin{figure}[bt]
    \centering
    \includegraphics[width=0.45\linewidth]{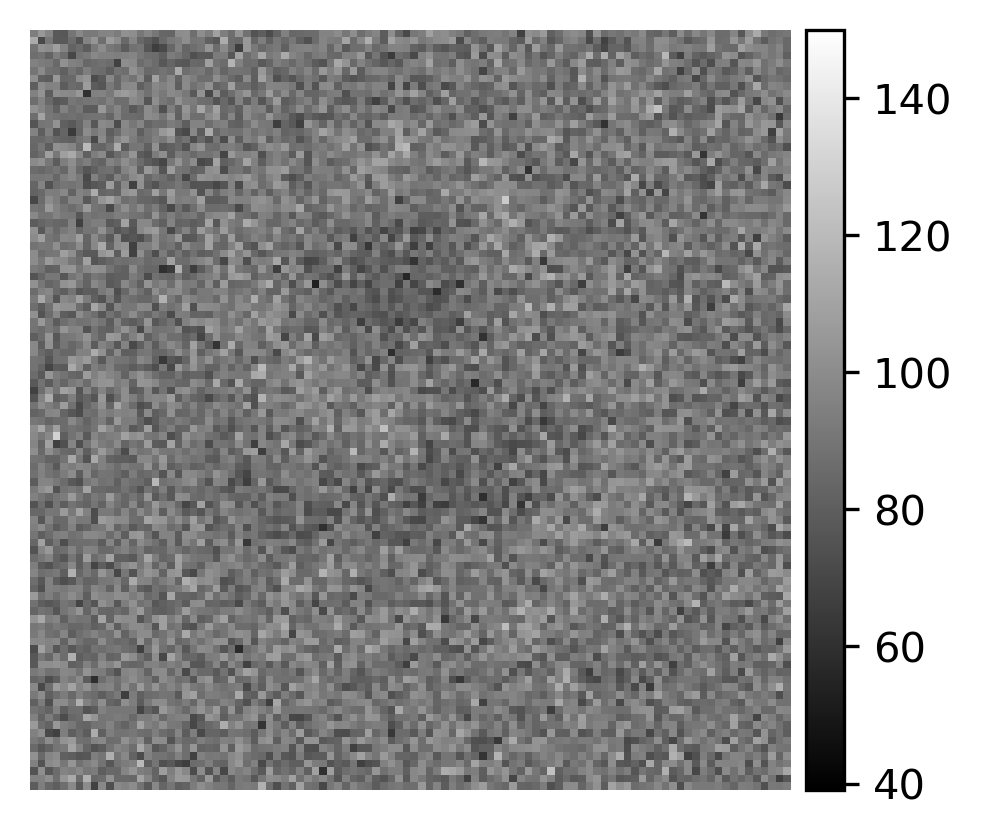}
    \hfill
    \includegraphics[width=0.45\linewidth]{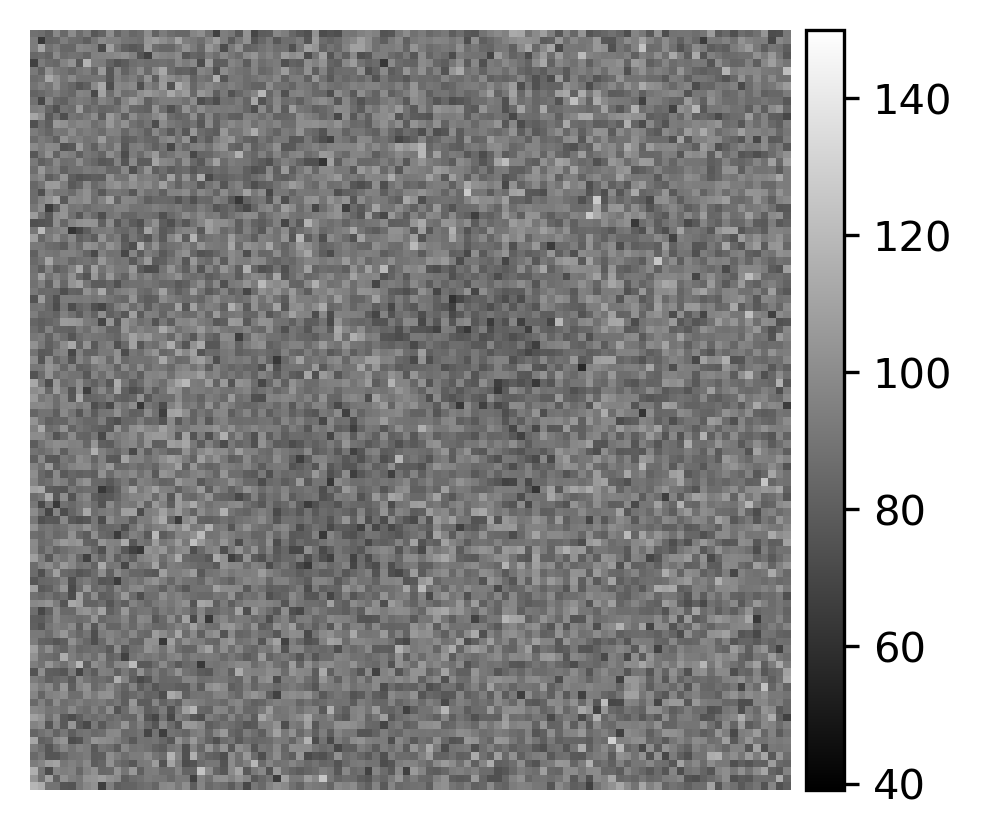}
    \caption{Example images generated from cryo-EM simulator Parakeet for the 4AKE dataset.}
    \label{fig:parakeet-images}
\end{figure}

We generated data through the use of the cryo-EM digital twin, Parakeet \citep{parkhurst_parakeet_2021}, which in turn relies on the MULTEM package \citep{lobato_multem_2015}.  MULTEM simulates electron--specimen interaction via high-order expansions of multislice solutions of the Schr\"{o}dinger equation and the frozen phonon model to simulate the effects of thermal diffuse scattering. 
That is, the generation of the data using the Parakeet package uses a different forward operator than our algorithms.
The Parakeet simulation also includes the frequency-dependent detector quantum efficiency, blurring due to microscopy optics modelled by the CTF, effects of partial spatial and temporal coherence, and Poisson noise that one typically observes in cryo-EM images. We used parameter settings (see Table~\ref{tab:parakeet_settings}) based on the Thermo Fisher Scientific Titan Krios microscope \citep{parkhurst_parakeet_2021}. Figure~\ref{fig:parakeet-images} shows examples of images used in our experiments. Observe that the effects of noise far outweigh the effect of the CTF. Data was generated using three different MD trajectories as ground truth. Details are provided in each experiment's relevant section.
   
\begin{table}
\centering 
\begin{tabular}{llll}
\toprule 
&\multicolumn{3}{c}{Datasets}\\
\cmidrule(lr){2-4}
Parameter  & 4AKE & BPTI & 5VZ0\\
\midrule
Electron Dose  & 90 $\text{e}^{-}$/$\textup{\AA}^{2}$  & 180 $\text{e}^{-}$/$\textup{\AA}^{2}$  & 90 $\text{e}^{-}$/$\textup{\AA}^{2}$ \\
Energy  & 300 keV  & 300 keV  & 300 keV \\
Defocus  & 2.0 $\mu$m  & 0.1 $\mu$m  & 3.0 $\mu$m \\
Spherical Aberration  & 2.7 mm  & 2.7 mm  & 2.7 mm \\
Chromatic Aberration  & 2.7 mm  & 2.7 mm  & 2.7 mm \\
Effective pixel size  & 0.1 nm  & 0.075 nm  & 0.2 nm \\
\bottomrule
\end{tabular}
\caption{Settings used for the Parakeet microscope simulator for the 4AKE, BPTI and 5VZ0 dataset. For a sensor with a pixel size of 14 $\mu$m pixels, this corresponds to $140,000\times$ magnification for the 4AKE dataset, $186,667\times$ for BPTI and $70,000\times$ for 5VZ0.}
\label{tab:parakeet_settings} 
\end{table}

In these experiments, as SGD is used in our algorithm, there is always an element of randomness in the estimated atomic positions. Hence, to ascertain the distribution of results that can be obtained by the model, the backbone was fitted to the same image data 200 times in order to obtain an estimate of the distribution. An SGD batch size of 10, 128 and 32 images for the 4AKE, BPTI and 5VZ0 datasets, was found to be good for introducing sufficient stochasticity to escape local minima yet gain a good fit to the data.

In this work the superimposed RMSD and TM scores are used to compare the estimated structures with the ground truth structure.
The superimposed RMSD between two conformations of the same protein, denoted $\text{RMSD}$, is the minimum RMSD over all rigid body transformations of the protein, where the RMSD is taken of the atomic positions of all the $C_\alpha$ atoms in the backbone. It is given by \citep{kufareva_methods_2012}
\[ \text{RMSD} (B,B') = \min_{Q \in \mathrm{SO}(3),\, \mathbf{d} \in \mathbb{R}^3} \sqrt{
\frac{1}{n} 
\sum_{i=0}^{n-1}
\left\|
\mathbf{r}_{3i + 1} 
- 
\left( Q \mathbf{r}_{3i + 1}' + \mathbf{d} \right)
\right\|_2^2
}\]

Measures like RMSD are sensitive to outliers -- one large deviation can have a disproportionate impact, and the RMSD is not normalised for protein length. For this reason, the TM score was developed \citep{zhang_scoring_2004, kufareva_methods_2012} and is defined by

\begin{equation*}
    \text{TM}(B,B') = \max_{Q \in \mathrm{SO}(3),\, \mathbf{d} \in \mathbb{R}^3} \left[
\frac{1}{n}
\sum_{i=0}^{n-1}
\frac{1}{1 + 
\left(
\frac{
\left\| 
\mathbf{r}_{3i+1} 
- \left( Q \mathbf{r}_{3i+1}' + \mathbf{d}
\right)
\right\|_2
}
{1.24 \cdot \sqrt[3]{n - 15} - 1.8}
\right)^2
}
\right]
\end{equation*}
The TM score is the maximum value (over all rotations and translations) of the mean of the transformed inverse squared distances between corresponding $C_{\alpha}$ atoms in the two protein structures (normalised for protein length). 
 Each penalty term in the sum minimally penalises small deviations, moderately penalises moderate deviations, and has a bounded penalty on large deviations. This gives a better measure of the global fold, and aligns more with human intuition \citep{zhang_scoring_2004, kufareva_methods_2012}.
 The normalisation term $1.24 \cdot \sqrt[3]{n - 15} - 1.8$, is the approximate average distance between residue pairs in randomly related proteins, obtained empirically from the $C_\alpha$ distances from 1,300 non-homologous PDB structures \citep{zhang_scoring_2004}, and normalises the measure for protein size, unlike the RMSD. A TM score of one indicates a perfect match and lower TM scores indicate poorer matches.

Comparison is made with a non-multiscale of the algorithm, which corresponds to \ref{alg:mscalealg}, with $K=2$ and $\sigma = 0$. We use a value of $K=2$ to show that restarting the optimiser at the current point (if momentum is used in SGD) does not cause the optimiser to find a more optimal point. 

\begin{table}[htbp]
\centering

\begin{tabular}{llll}
\toprule
Parameter / Attribute & 4AKE & BPTI & 5VZ0 \\
\midrule
Multiscale blur, $\sigma$ ($\mathring{\mathrm{A}}$)   & 4           & 2           & 5           \\
Multiscale levels, $K$                               & 5           & 5           & 5           \\
Stop criterion (abs. tolerance)                      & $10^{-5}$   & $10^{-5}$   & $10^{-5}$   \\
Batch Size                                          & 10          & 128         & 32          \\
\bottomrule

\end{tabular}

\caption{Parameters and attributes used for the multiscale algorithm for 4AKE, BPTI, and 5VZ0.}
\label{tab:experiment_params}
\end{table}

The details of the multiscale algorithm parameters used for each dataset can be found in Table~\ref{tab:experiment_params}.

\subsection{4AKE results}

\begin{figure*}[htbp!] 
\centering

\begin{subfigure}[b]{0.47\linewidth}
  \includegraphics[width=\linewidth,height=0.33\textheight,keepaspectratio]{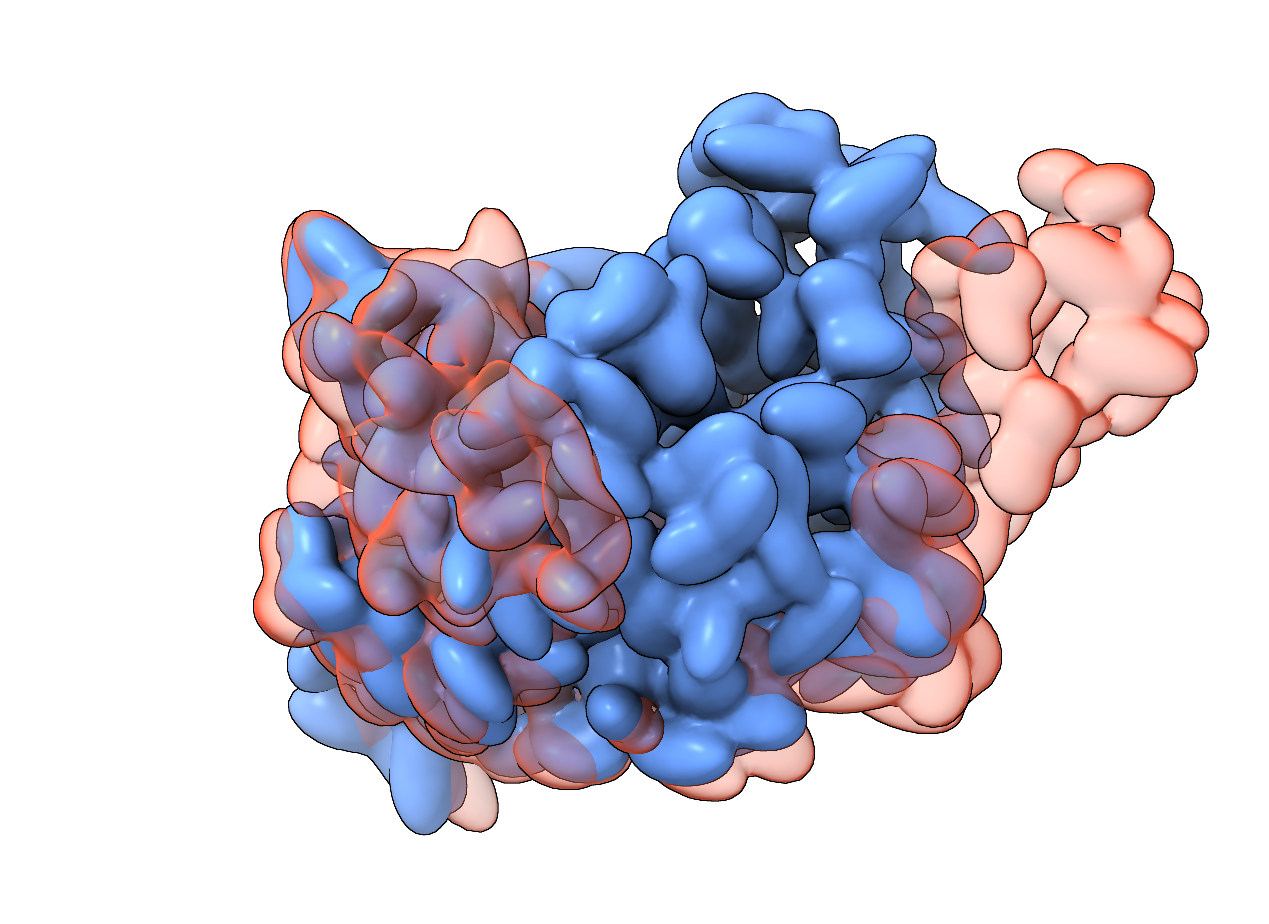}
  \caption{Template (blue) and target (red) conformation visualisation for the 4AKE dataset.}
  \label{fig:temp-targ-4AKE}
\end{subfigure}\hfill
\begin{subfigure}[b]{0.47\linewidth}
  \includegraphics[width=\linewidth,height=0.33\textheight,keepaspectratio]{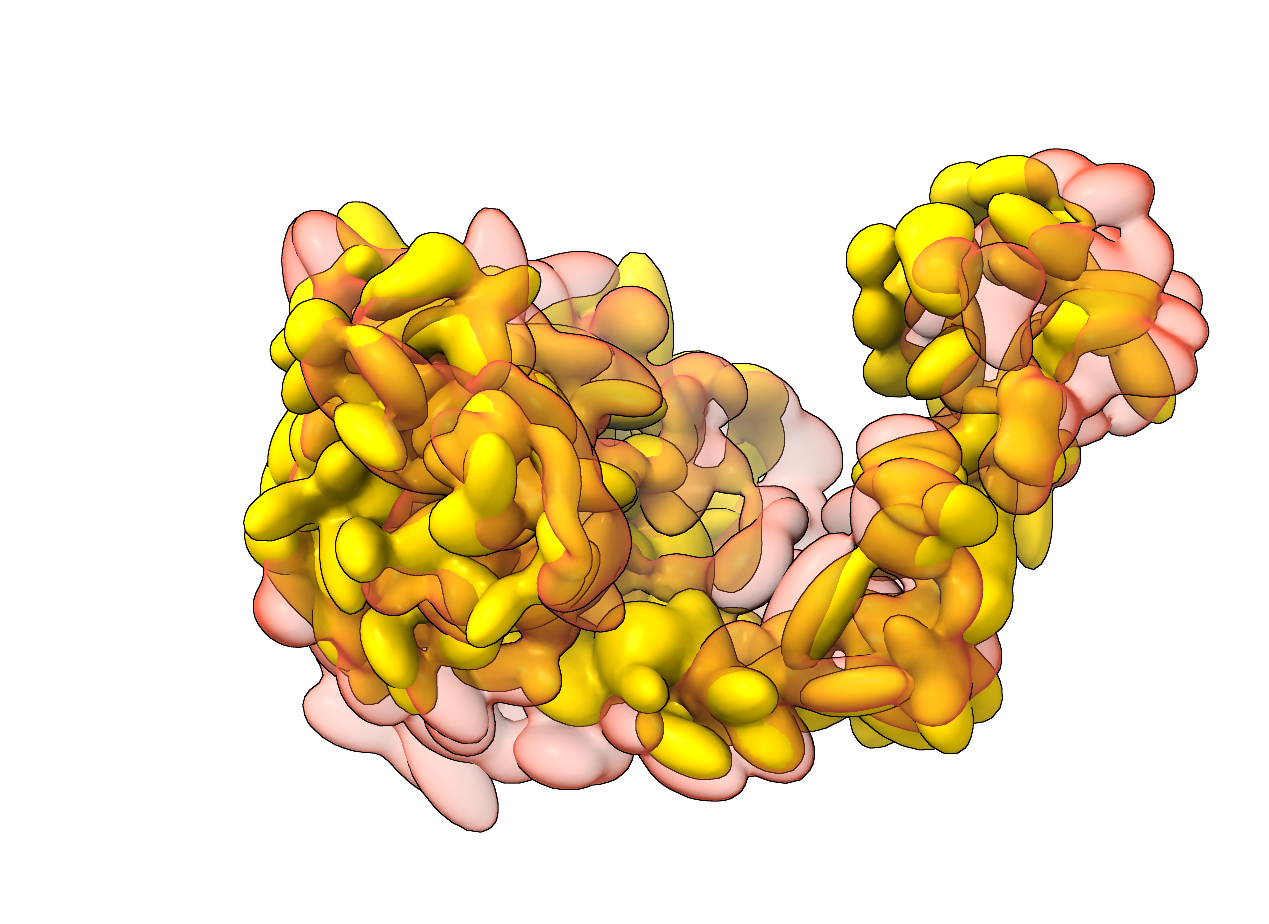}
  \caption{Example fitted density using multiscale approach. Estimated density is yellow, target density is red.}
  \label{fig:ex-fitted}
\end{subfigure}

\par\medskip 

\begin{subfigure}[b]{0.47\linewidth}
  \includegraphics[width=\linewidth,height=0.33\textheight,keepaspectratio]{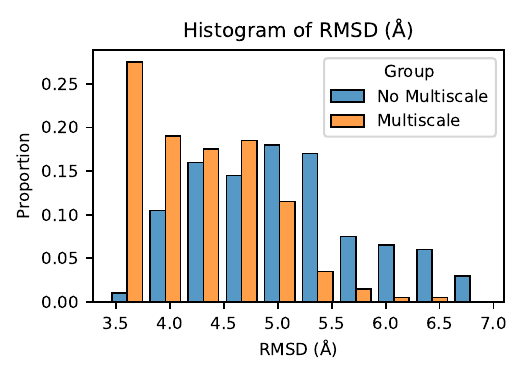}
  \caption{RMSD histogram comparing multiscale and non-multiscale algorithms. RMSD of target vs. template conformation is 6.8 \AA.}
  \label{fig:rmsd}
\end{subfigure} \hfill
\begin{subfigure}[b]{0.47\linewidth}
  \includegraphics[width=\linewidth,height=0.33\textheight,keepaspectratio]{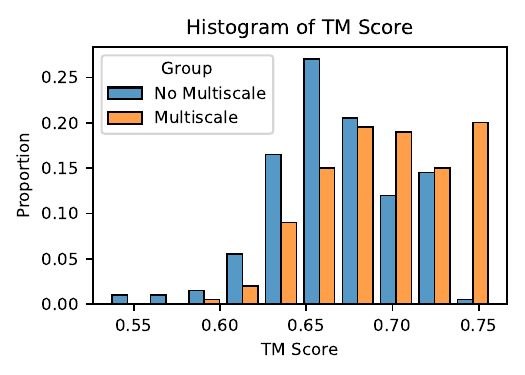}
  \caption{TM score histogram comparing multiscale and non-multiscale algorithms. TM score of target vs. template conformation is 0.69.}
  \label{fig:tm-score}
\end{subfigure}

\par\medskip

\begin{subfigure}[b]{0.32\linewidth}
  \includegraphics[width=\linewidth,height=0.20\textheight,keepaspectratio]{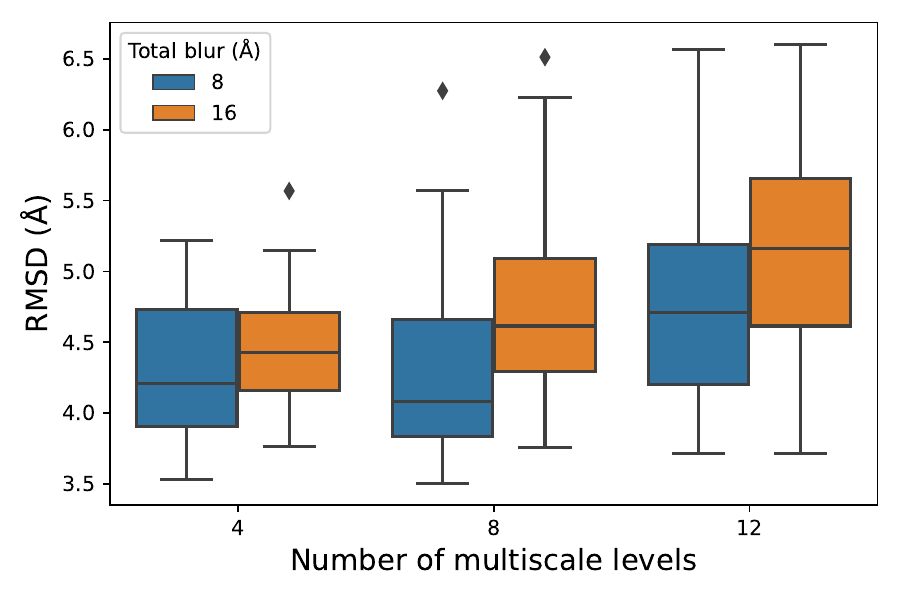}
  \caption{Boxplots of RMSDs obtained from computer runs with different numbers of multiscale levels, for the 4AKE dataset.}
  \label{fig:sweep_multiscale_blur}
\end{subfigure}
    \begin{subfigure}[b]{0.32\linewidth}
        \includegraphics[width=\linewidth]{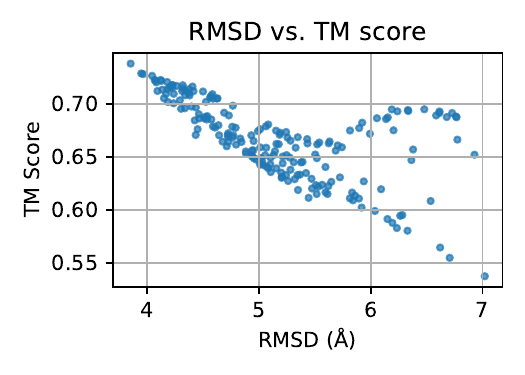}
        \caption{TM score vs. RMSD, without multiscale.} \label{fig:rmsd-tm-4ake-1}
    \end{subfigure}
    \begin{subfigure}[b]{0.32\linewidth}
        \includegraphics[width=\linewidth]{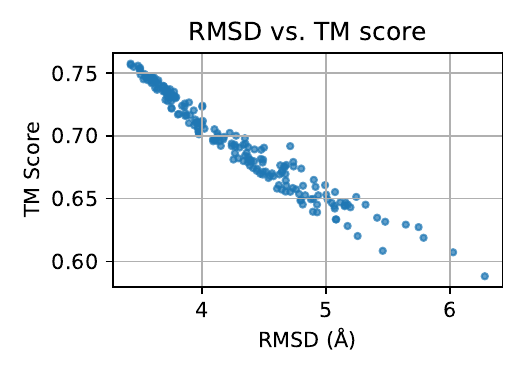}
        \caption{TM score vs. RMSD, with multiscale.} \label{fig:rmsd-tm-4ake-2}
    \end{subfigure}

\caption{Visualisations for the 4AKE dataset.}
\label{fig:combined-4AKE}
\end{figure*}

Simulated cryo-EM data was generated using a molecular dynamics trajectory of the apo-adenylate kinase of E. coli, also known as 4AKE, moving from the closed state to the open state \citep{beckstein_simulated_2018}, comprised of $5,000$ particle images and pose angles. The target and template were chosen from the molecular dynamics simulation to be the open and closed states of the molecule (see Figure~\ref{fig:temp-targ-4AKE}).

Figure~\ref{fig:temp-targ-4AKE} shows the densities of the template and target conformations for the 4AKE dataset. There are many small-scale differences, but the most notable difference seen is the opening of the protein on the right hand side. Figure~\ref{fig:ex-fitted} shows an example fitted atomic structure produced by our method. Observe that the method has been able to successfully model the opening of the conformation, as seen when the estimated and true densities are superimposed  on each other.

As can be seen from Figures \ref{fig:rmsd} and \ref{fig:tm-score} 
there is a substantial improvement with respect to the non-multiscale version of the algorithm, in both the RMSDs and TM scores of the fitted backbone versus the true backbone. In the TM score plot, observe that there appears to be a clustering of results at a TM score of 0.75, indicating there is a tendency for the algorithm to avoid large deviations from the true structure.

Another question that can be asked is whether any additional benefit can be yielded through increasing the final blur level or increasing the number of multiscale levels. We performed 400 computer runs, randomising over two levels of blur (8 and 16 Å) and three different settings of multiscale levels (4, 8, 12 levels). As can be seen from the results in Figure~\ref{fig:sweep_multiscale_blur}, a slight improvement can be obtained by increasing the number of multiscale levels to 8, keeping the final blur at 4 Å, but no improvement is obtained from adding further blur, or by increasing the number of multiscale levels to 12.

An interesting phenomenon can be observed when the TM scores are plotted against RMSD. As mentioned earlier, the RMSD is dominated by large deviations, but TM score penalises moderate deviations the most, which in this case corresponds to deviations around $3.14$ Å for the 4AKE molecule.

Figures~\ref{fig:rmsd-tm-4ake-1} and \ref{fig:rmsd-tm-4ake-2} show plots of the RMSD versus the TM score, where we observe that as RMSD decreases (an improvement), TM score increases (an improvement). Yet, in the runs without multiscale, there are runs that produce high values of RMSD, indicating poor fit, but also high values of TM score. An explanation is that, since the RMSD is dominated by large deviations, there are large misalignments, but as the TM score is dominated by small to medium deviations, and large deviations have a bounded penalty, then, this indicates that whilst many of the deviations are small- to medium-sized, some large deviations remain after fitting with the non-multiscale algorithm. These do not occur, as can be seen, on the plot of the multiscale results. This occurs not just for RMSDs between $5$ and $7$ Å but also at smaller values of RMSD in the non-multiscale runs, but are not present in the multiscale runs.
This could be seen as evidence that the multiscale algorithm is allowing the alignment of long range deviations not possible in the non-multiscale algorithm. Through coarsening by convolving with a Gaussian, this ``spreads'' out the density over a larger volume, allowing longer-range deviations to be detected and aligned.

\subsection{BPTI results}

\begin{figure}[htbp!]
    \centering
\begin{subfigure}[b]{\linewidth}
    \includegraphics[width=0.45\linewidth]{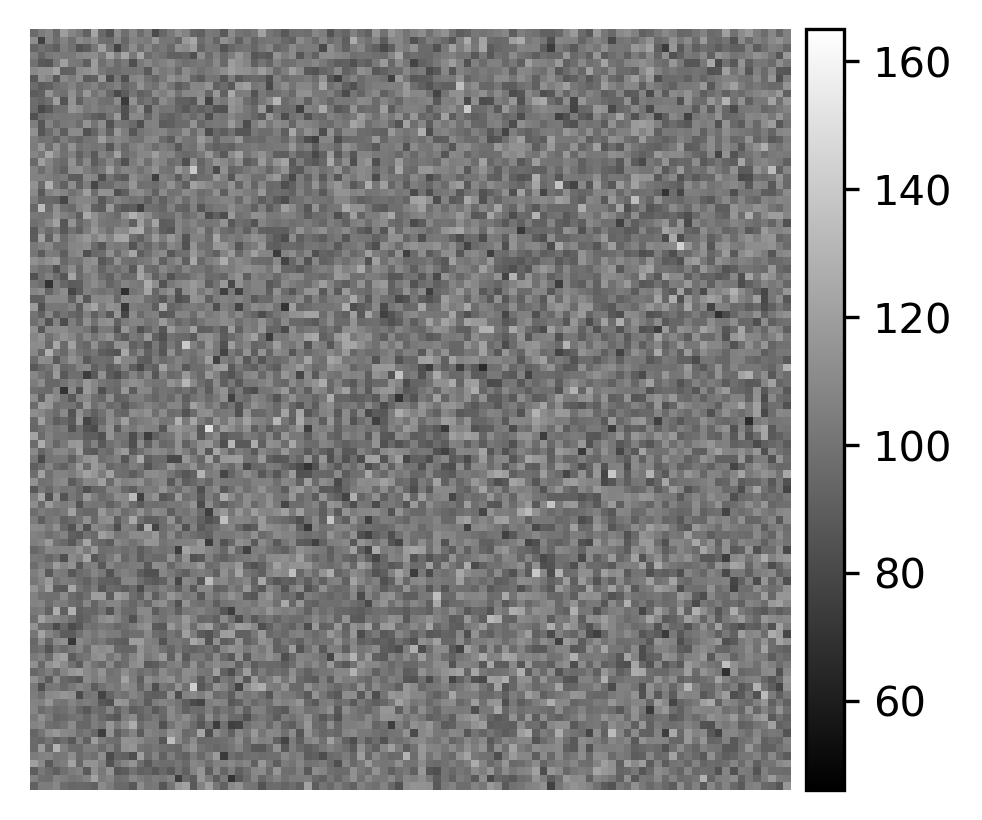}
    \hfill
    \includegraphics[width=0.45\linewidth]{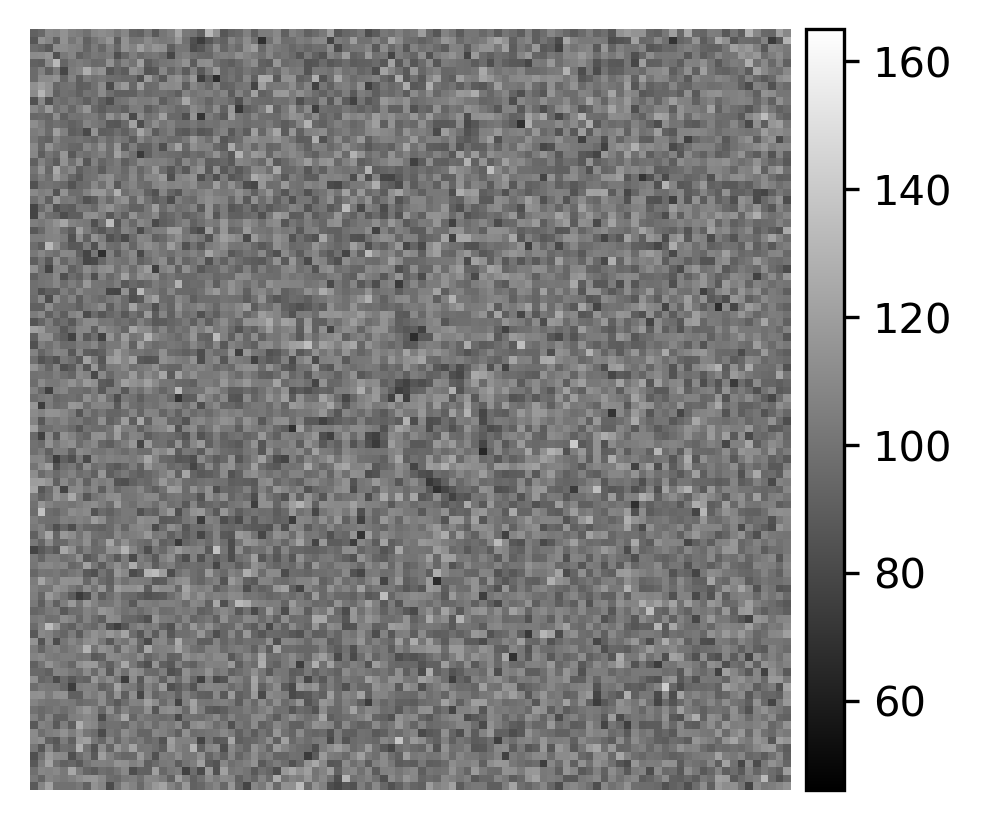}
    \caption{Two random particle images from the BPTI dataset.}
    \label{fig:bpti-image}
\end{subfigure}

\begin{subfigure}[b]{\linewidth}
    \includegraphics[width=0.45\linewidth]{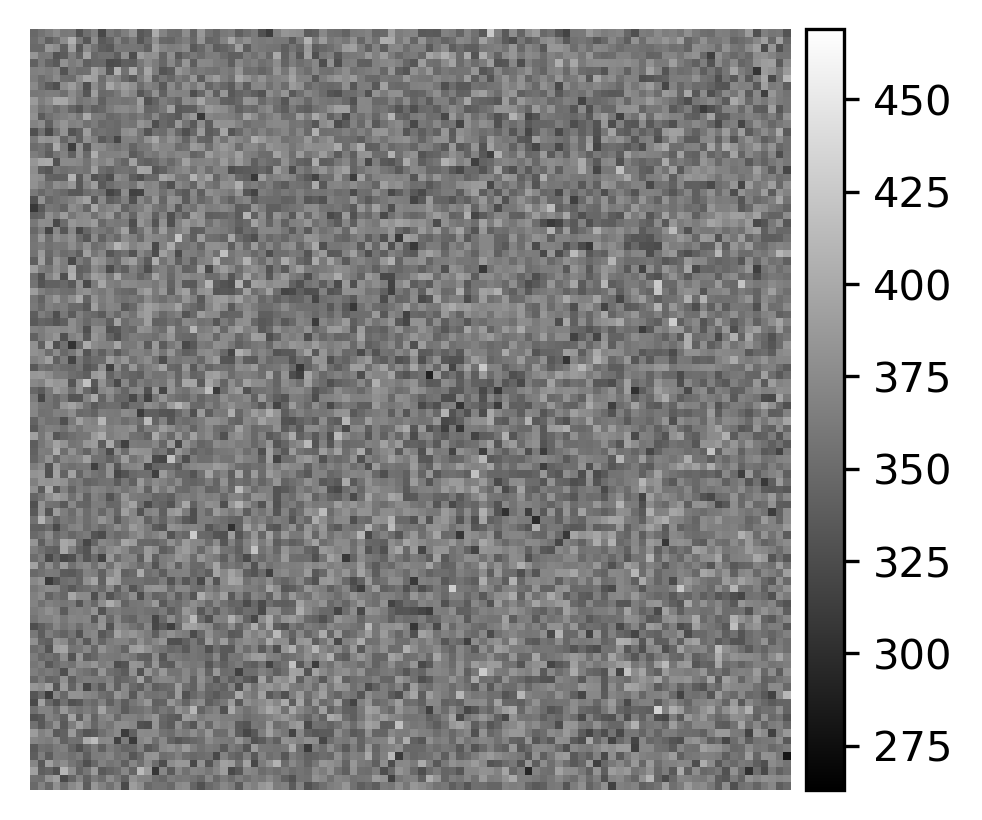}
    \hfill
    \includegraphics[width=0.45\linewidth]{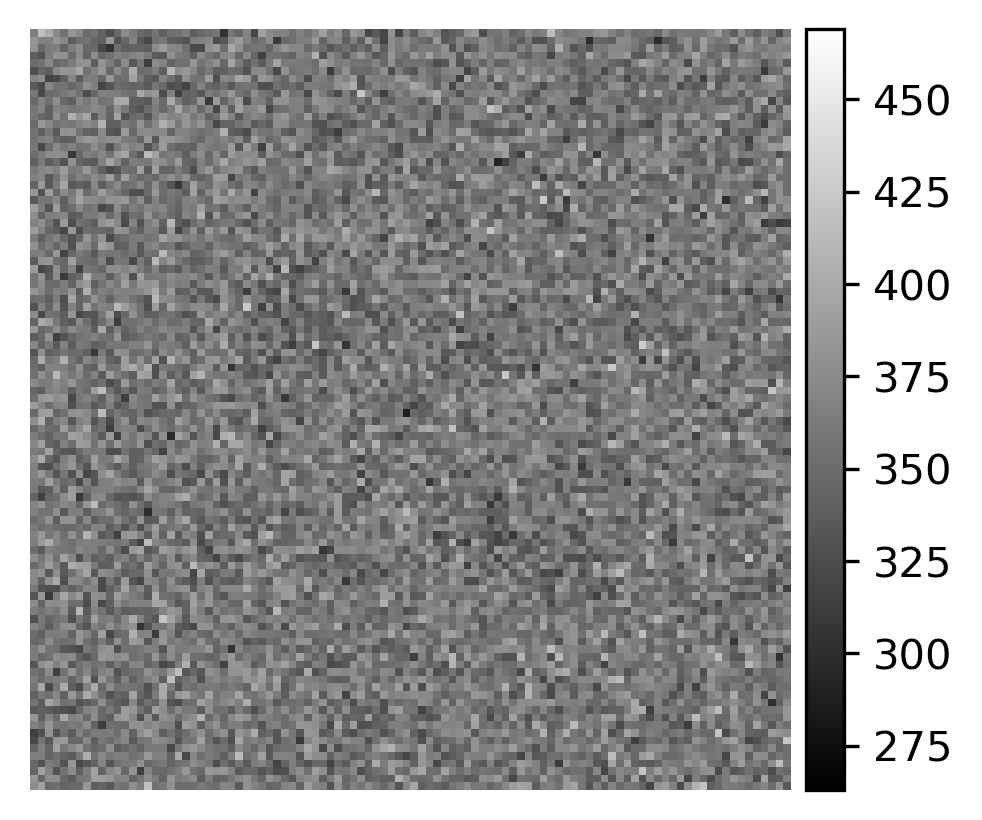}
    \caption{Two random particle images from the 5VZ0 dataset.}
    \label{fig:5VZ0-image}
\end{subfigure}

    \caption{Example particle images from the BPTI and 5VZ0 datasets. The bar on the right is the electron count.}
\end{figure}

The BPTI dataset was more challenging as its particle size was much smaller than for the 4AKE dataset. The effective pixel size was set to 0.75 \AA, the defocus was set to $0.1$~{\textmu}m (to make sure enough fine detail was visible), and the electron dose was set to $180 \, \text{e}^{-}/\textup{\AA}^2$ (to compensate for low contrast caused by the low defocus). For a sensor with a pixel size of 14 {\textmu}m pixels, this corresponds to $186,667\times$ magnification. Still, despite these settings, the particle (see Figure~\ref{fig:bpti-image}) is barely visible. As with the 4AKE dataset, $5,000$ particle images and their pose angles were generated with the digital twin electron microscope.

\begin{figure*}[htbp!]
    \centering

    \begin{subfigure}[b]{0.48\linewidth}       \includegraphics[width=\linewidth,height=0.33\textheight,keepaspectratio]{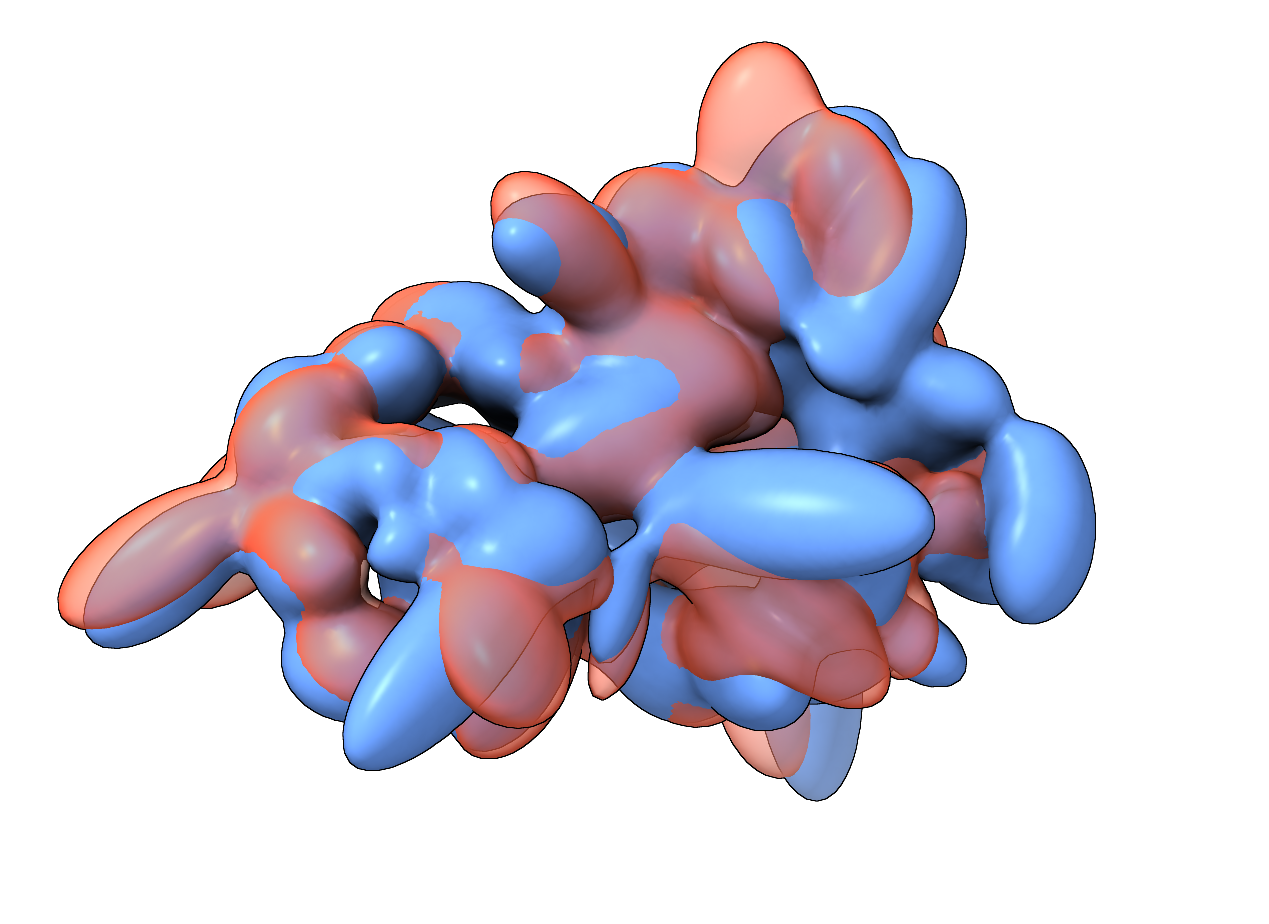}
        \caption{Template (blue) and target (red) conformation visualisation for the BPTI dataset.}
        \label{fig:temp-targ-BPTI}
    \end{subfigure}
    \hfill
    \begin{subfigure}[b]{0.48\linewidth}
        \includegraphics[width=\linewidth,height=0.33\textheight,keepaspectratio]{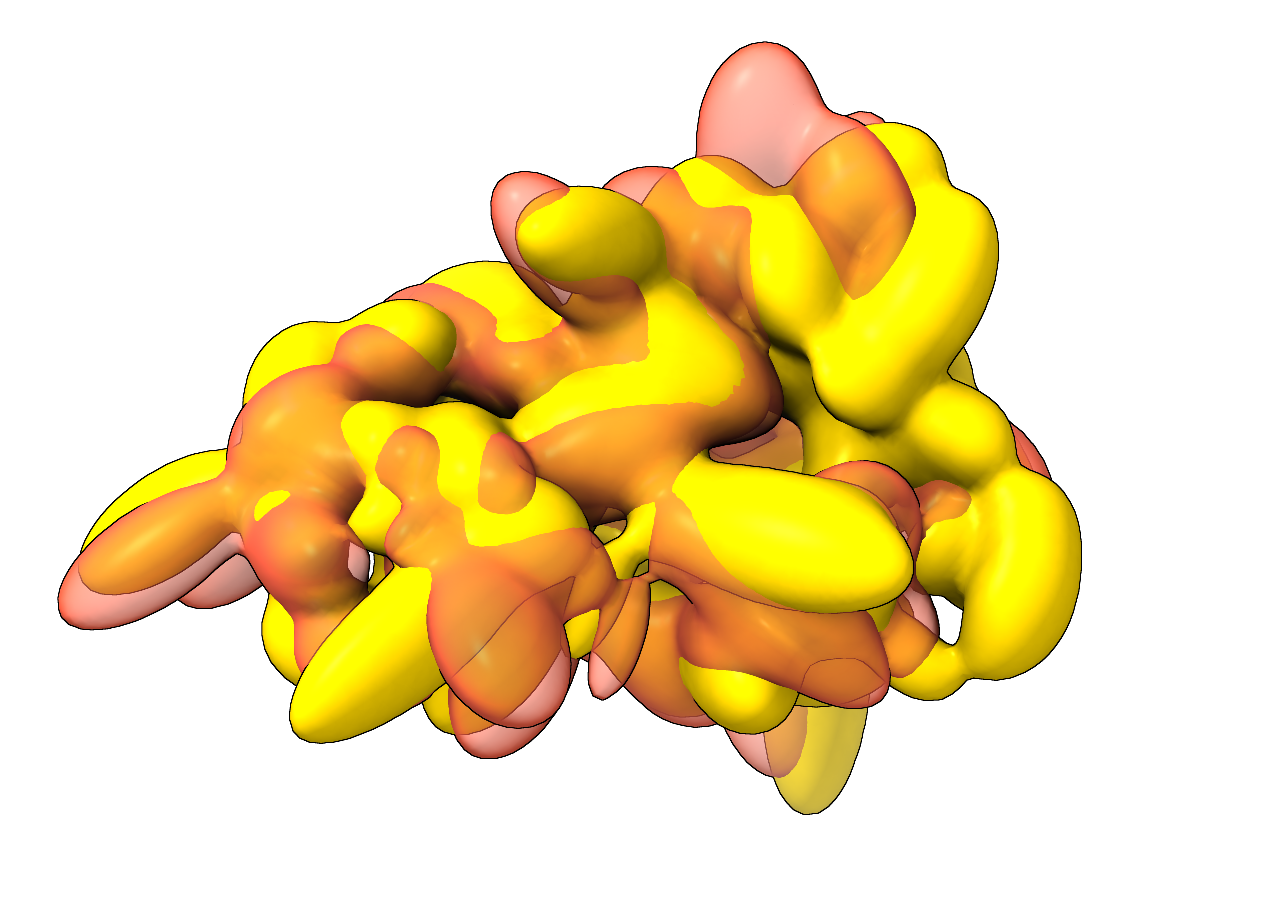}
        \caption{Example fitted density using multiscale approach. Estimated density is yellow, target density is red.}
        \label{fig:ex-fitted-BPTI}
    \end{subfigure}

\vspace{5em}

\begin{subfigure}[b]{0.48\linewidth}
    \includegraphics[width=\linewidth]{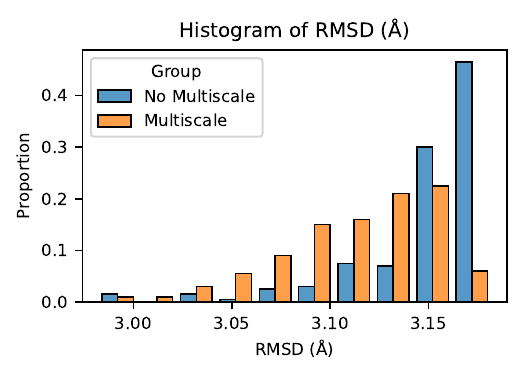}
    \caption{RMSD histogram comparing multiscale and non-multiscale algorithms for BPTI. RMSD of target vs. template conformation is 3.1 \AA.}
    \label{fig:hist-RMSD-with-without-BPTI}
\end{subfigure} 
\hfill
\begin{subfigure}[b]{0.48\linewidth}
    \includegraphics[width=\linewidth]{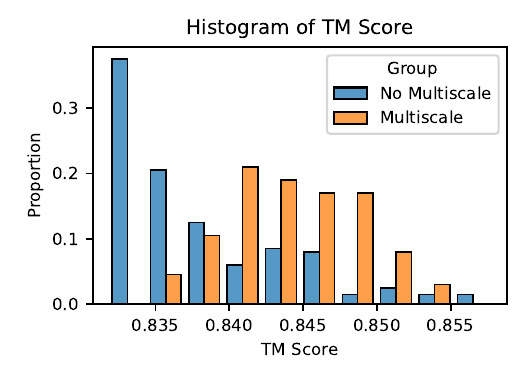}
    \caption{TM score histogram comparing multiscale and non-multiscale algorithms for BPTI. TM score of target vs. template conformation is 0.83.}
    \label{fig:hist-TM-with-without-BPTI}
\end{subfigure}

\caption{Visualisations for the BPTI dataset.}
\label{fig:combined-BPTI}

\end{figure*}

As can be seen in Figure~\ref{fig:hist-RMSD-with-without-BPTI}, the use of the multiscale approach results in smaller RMSDs, indicating improved model fit. This improvement is more clearly seen in the TM scores (Figure~\ref{fig:hist-TM-with-without-BPTI}). This is thus evidence that the multiscale algorithm is able to improve global fit and lessen the spread of results obtained from SGD. The target and template densities are shown in Figure~\ref{fig:temp-targ-BPTI} and an example fitted density is shown in Figure~\ref{fig:ex-fitted-BPTI}.

\subsection{5VZ0 results}

\begin{figure*}[htbp!]
    \centering

    \begin{subfigure}[b]{0.48\linewidth}       \includegraphics[width=\linewidth,height=0.33\textheight,keepaspectratio]{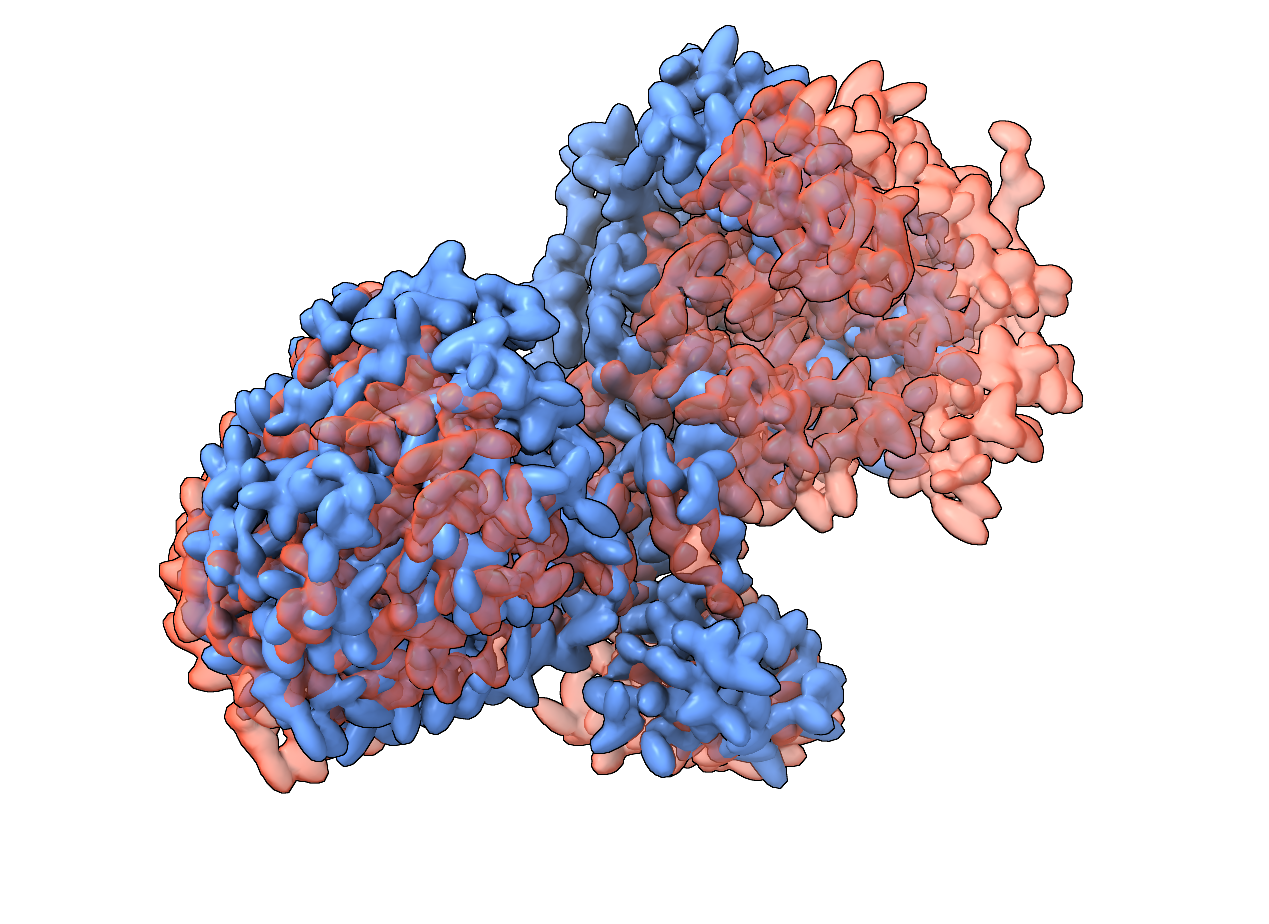}
        \caption{Template (blue) and target (red) conformation visualisation for the 5VZ0 dataset.}
        \label{fig:temp-targ-5VZ0}
    \end{subfigure}
    \hfill
    \begin{subfigure}[b]{0.48\linewidth}
        \includegraphics[width=\linewidth,height=0.33\textheight,keepaspectratio]{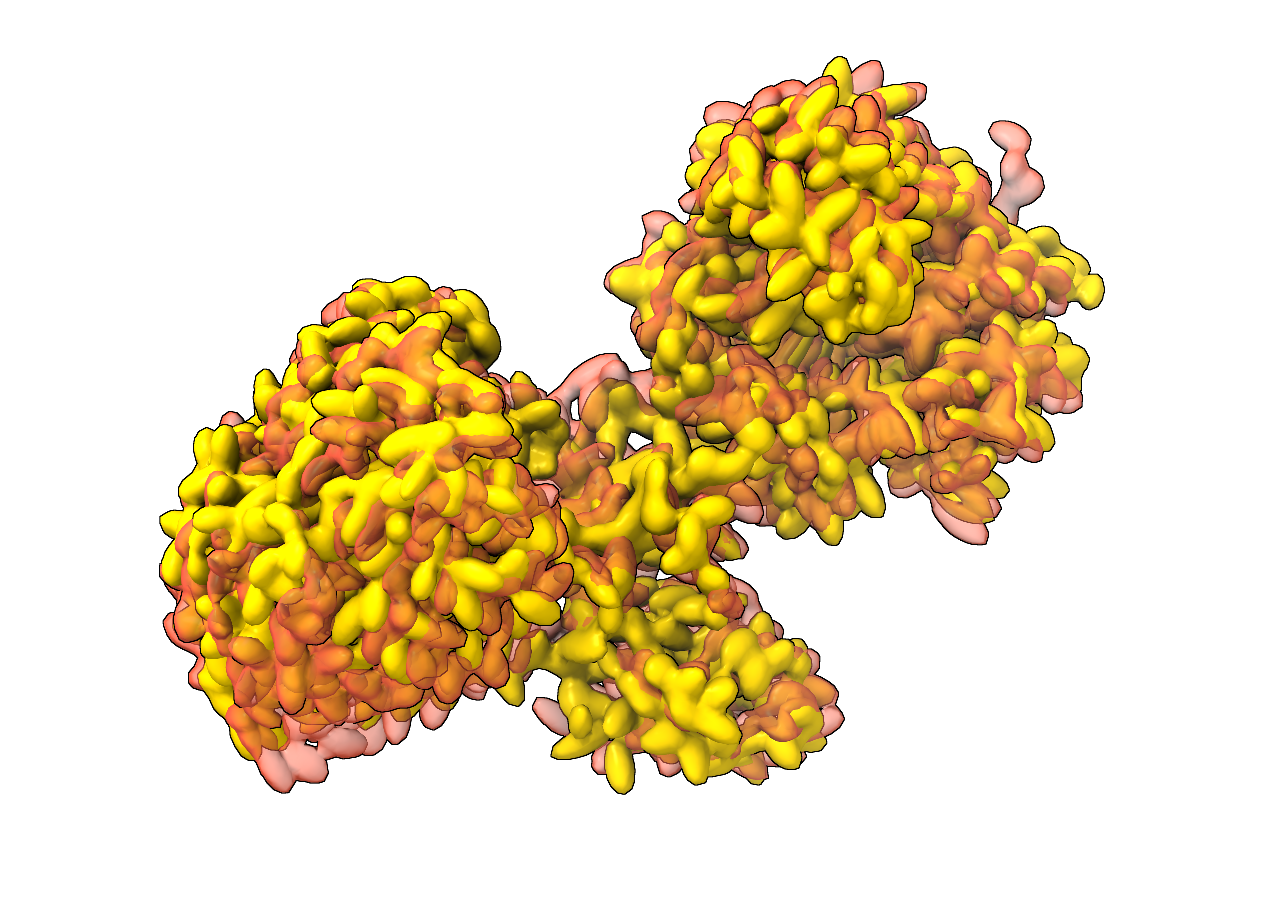}
        \caption{Example fitted density using multiscale approach. Estimated density is yellow, target density is red.}
        \label{fig:ex-fitted-5VZ0}
    \end{subfigure}

\vspace{5em} 

\begin{subfigure}[b]{0.48\linewidth}
    \includegraphics[width=\linewidth,height=0.33\textheight,keepaspectratio]{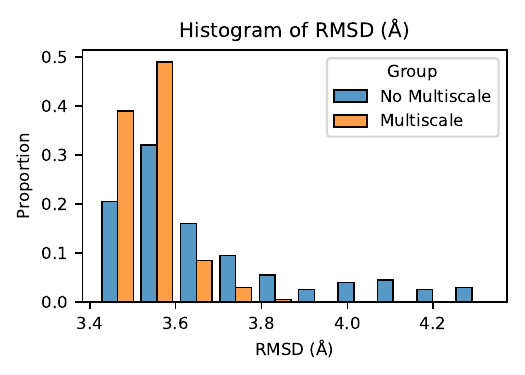}
    \caption{RMSD histogram comparing multiscale and non-multiscale algorithms for 5VZ0. RMSD of target vs. template conformation is 4.4 \AA.}
    \label{fig:hist-RMSD-with-without-5VZ0}
\end{subfigure}
\hfill
\begin{subfigure}[b]{0.48\linewidth}
    \includegraphics[width=\linewidth,height=0.33\textheight,keepaspectratio]{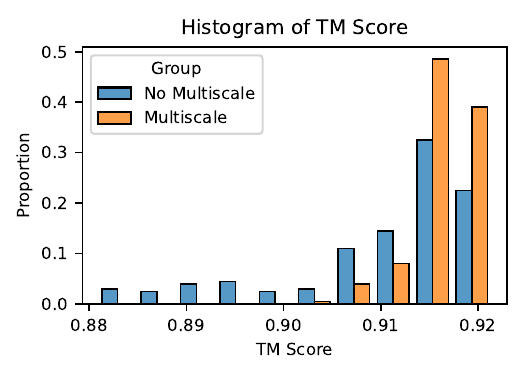}
    \caption{TM score histogram comparing multiscale and non-multiscale algorithms for 5VZ0.  TM Score of target vs. template conformation is 0.88.}
    \label{fig:hist-TM-with-without-5VZ0}
\end{subfigure} 

\caption{Visualisations for the 5VZ0 dataset. Top row: template and target conformation visualisation (left) and fitted backbone structure (right). Bottom row: RMSD histogram (left) and TM score histogram (right) comparing multiscale versus non-multiscale algorithms.}
\label{fig:combined-5VZ0} 

\end{figure*}

The 5VZ0 dataset (of a single chain of the crystal structure of Lactococcus lactis pyruvate carboxylase G746A mutant in complex with cyclic-di-AMP) is based off a protein trajectory stored with the ATLAS project \citep{vandermeersche_atlas_2024}. This MD trajectory models single protein chain, which consists of 1,144 residues. We generated simulated data from this snapshot using Parakeet to create 5,000 images of the protein under random pose angles. In this dataset, the electron dose was set to 90 $\text{e}^{-}/\mathring{\mathrm{A}}^{2}$. The molecule is large so an effective pixel size of $2\ \mathring{\mathrm{A}}$ and defocus of $3\ \mu m$ is used. For a sensor with a pixel size of 14 {\textmu}m pixels, this corresponds to $70,000\times$ magnification. Again, the noise level is high enough that the molecule is barely visible from individual images (Figure  \ref{fig:5VZ0-image}).
Figure~\ref{fig:temp-targ-5VZ0} shows the densities of the template and target conformations of the molecule. Figure~\ref{fig:ex-fitted-5VZ0} is an example density fitted by the multiscale algorithm.

As with the previous datasets, use of the multiscale algorithm produces a skew in the distribution of the RMSDs to lower values and TM scores to higher values, indicating better quality of fit over the non-multiscale algorithms (see Figures \ref{fig:hist-RMSD-with-without-5VZ0} and \ref{fig:hist-TM-with-without-5VZ0}).

\section{Discussion}

\begin{figure}[ht]
    \centering
    \includegraphics[width=\linewidth]{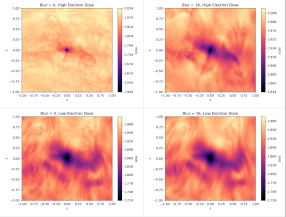}
    \caption{Plots of 2D slices of the loss surface on the same planes, at the same position in high dimensional space, visualised using the method of \citet{li_visualizing_2018}. ``High'' election dose is $6\times10^{11}\,e^{-}/\text{\AA}$, ``low'' dose is $90\,e^{-}/\text{\AA}$}
    \label{fig:vis-loss-surf}
\end{figure}

It is clear that the multiscale approach proposed here yields substantial advantages over a non-multiscale approach.  Multiscale approaches have been largely absent from joint reconstruction and model-building methods. Each of \citet{rosenbaum_inferring_2021}, \citet{chen_improving_2024} and \citet{li_cryostar_2024} involve fitting a coarse-grained atomic model to the data but involve no iterative refinement. 

There are several possible explanations of why the multiscale approach yields this benefit. One possible explanation is that a coarse-to-fine approach encourages large-scale features to be aligned first, before smaller features. The coarsening through blurring is an attenuation of high-frequency Fourier coefficients or a removal of fine detail. A related approach has been used in RELION \citep{scheres_relion_2012} for two-stage angular refinement: sampling over a coarse angular grid, then refining over a higher-resolution grid. Sampling over a coarse grid forces alignment of major features, and the fine grid refines finer features. Similarly, CryoSTAR \citep{li_cryostar_2024} performs a low-pass filter on the input data, so only low-frequency information is used, but there is no subsequent refinement phase. This does indicate that the majority of the information which can be used to perform reconstruction and model fitting is of low spatial frequency.

Another perspective on why multiscale is able to lessen the impact of non-convexity is that it smoothens the loss surface. The loss surface is difficult to visualise, as it is high-dimensional. However, using the method of \citet{li_visualizing_2018}, it is possible to view 2D slices of the loss function (see Figure~\ref{fig:vis-loss-surf}). The left shows the loss function when no blur is applied to the Gaussians and the right shows the loss function when the Gaussians are convolved with a Gaussian blur of 16.0 \AA{}.

We can observe that the loss surface is rough, and that coarsening removes some of the roughness, which may cause first-order gradient-based optimisers to converge to poor local minima. Even though this smoothing looks slight, it is more significant than meets the eye, as in high-dimensional space, most directions are orthogonal, so only a small distance is traversed in each plane. Furthermore, the top row of the diagram shows the effect of coarsening for a very high electron dose (the high SNR case) while the bottom row shows a low electron dose (the low SNR case). In the top row, clearly small minima are removed and the loss surface is smoothed (note that the colour scale on the right hand plot goes between a smaller range of values). In the bottom row, small minima are removed, but the high noise level introduces many minima and valleys which do not seem to be removed by this type of coarse-graining. We have investigated other methods of coarsening, including locking bond angles, but these did not yield any significant improvement.

Yet another possible reason why multiscale is able allow convergence to a better local minima is that our form of coarsening alters the loss surface and the optimiser path. Coarsening the representation of the density changes the loss function, altering the optimisation landscape (smoothing it in certain areas), and encourages the optimisation algorithm to take another path. Smoothing the loss landscape prevents the optimisation algorithm from converging to shallow and narrow local minima, favouring wider or deeper local minima.

\begin{figure}[tb]
    \centering
    \includegraphics[width=1.0\linewidth]{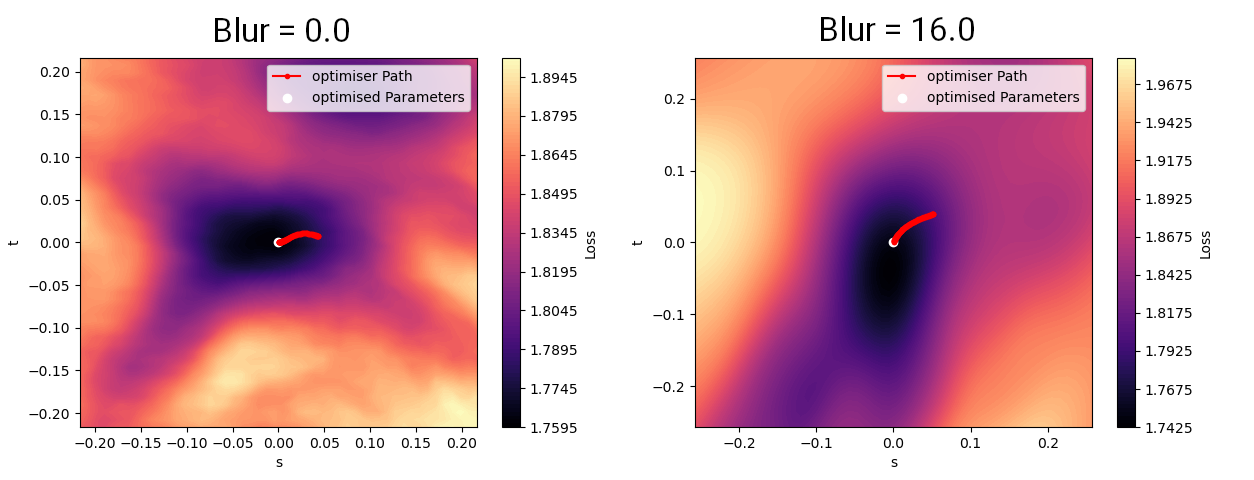}
    \caption{Plots of 2D slices of the loss surface along the first two principal components of the optimiser paths, using the method of \citet{li_visualizing_2018}}
    \label{fig:opt-path}
\end{figure}

An illustration of how our form of coarsening alters the loss function is as follows. Most of the mass in a Gaussian dictionary element is concentrated in a small volume. Consider a toy example, where part of a chain is represented by an isometric Gaussian with width parameter $\sigma$. The majority of the mass is within $2\sigma$ of its centre. As a result, if that part of the chain were misaligned more than $2\sigma$, with the $L^2$ loss, movement of that part of the chain would yield no improvement in loss. However, if $\sigma$ were increased sufficiently, this would not be the case. Figure~\ref{fig:opt-path} shows that two different optimiser paths are take to two different local minima, and that the loss surface is visibly smoother in the coarsened case. 

The algorithm is also fast compared to other algorithms that have been previously developed with a mean runtime of $421$~s for 4AKE on a single NVIDIA A40 GPU. In comparison, \citet{rosenbaum_inferring_2021} required several days to converge, using eight TPUv3 devices.

The multiscale approach is capable of performing joint reconstruction and model-building to high accuracy, and shows robustness to high levels of noise and CTF. Since in cryo-EM, the ground truth structure is unknown in real-world datasets, accurate validation presents its own challenge. Previous works evaluate their methods using synthetic datasets, for example,  
\citet{rosenbaum_inferring_2021} use ten times the electron dose (compared to what would be typically seen in single-particle cryo-EM) in their simulation studies, with 0.5~{\textmu}m defocus, \citet{zhong_exploring_2021} tested on a synthetic dataset without CTF and Gaussian noise at an SNR of 0.1. CryoSTAR, to validate their algorithm, perform a test of their algorithm on a synthetic data set generated with Gaussian noise with a SNR of 0.0001. To obtain a more faithful evaluation, we instead use a electron microscope digital twin, which is independent from the forward model used in our method for joint reconstruction and model building. This shows that our algorithm is able to retrieve the ground truth structure from a physically realistic scenario and not a simplified model. Through testing our algorithms with a separate microscope digital twin, we have shown the robustness of the forward model under realistic conditions. 

Explicitly modelling the bonding and backbone structure increases the likelihood that the atomic models obtained are reasonable. Other algorithms use penalty terms in the loss function to encourage solutions that preserve structure, but they do not guarantee that bonds will be preserved. For example, \citet{rosenbaum_inferring_2021}, \citet{zhong_exploring_2021}, and \citet{li_cryostar_2024} use two or three different penalty terms in the loss function to enforce structural constraints. These additional penalty terms introduce more tuning parameters into the model and increase computation time. Using penalty terms as structural priors also does not ensure that all structures obtained from such methods may make physical sense. For example, in the Cryofold paper \citep{zhong_exploring_2021}, the authors report that during experiments with homogenous reconstruction, they saw different types of misfits, such as instances where several different RBFs (representing different backbone atoms or side chains), were placed at the same spatial position. Explicitly modelling the bonding and backbone structure lowers the likelihood of these types of problems as the chain representation limits the possibility of conformations. Looking forward, there are several different routes in which the algorithm can be extended to heterogenous reconstruction, for example using a VAE to model continuous heterogeneity as in, for example, \citet{rosenbaum_inferring_2021}, \citet{zhong_exploring_2021}, and \citet{li_cryostar_2024}.

The algorithm is also easily extendable to multiple protein chains. In this case optimisation would be carried over all chains, and after the global alignment algorithm, a whole-chain alignment phase can be inserted, where each chain is adjusted as a rigid segment, before optimisation over torsion and bond angles.

\section{Conclusion}

We have developed a new multiscale algorithm for the simultaneous reconstruction and model fitting to cryo-EM image data. Through the use of a strong prior induced by the parametrisation of the protein backbone by bonds and bond angles, we greatly reduce the ill-posedness of the problem, whilst keeping computational complexity low (as not every atom-to-atom interaction needs to be modelled). We use a computationally efficient forward operator approximation, bypassing the effect of CTF to reduce computational costs. Through the use of multiscale iterative refinement, we also reduce the likelihood of the optimisation process converging to poor local minima. We believe that this technique will be of great benefit for all researchers using cryo-EM as a tool for protein reconstruction.

\section*{Funding}

This work was supported by the Knut and Alice Wallenberg (KAW) Foundation through a joint WASP--DDLS grant.

\section*{Acknowledgements}

We would like to thank Erik Lindahl for fruitful discussions and helpful suggestions.

\section*{Declaration of competing interest}
The authors declare that they have no known competing financial interests or personal relationships that could have influenced the work reported in this paper.

\section*{Data availability}
Simulated cryo-EM data and analysis code used in this study are available at [Repository URL/DOI]. 

\section*{CRediT authorship contribution statement}

\noindent \textbf{David Y. W. Thong:} Conceptualization, Methodology, Software, Validation, Writing - original draft, Writing - review \& editing \\
\textbf{Ozan \"{O}ktem:} Conceptualization, Methodology, Project administration, Supervision, Funding acquisition, Writing - review \\
\textbf{Joakim And\'{e}n:} Conceptualization, Methodology, Validation, Project administration, Funding acquisition, Supervision, Writing - review \& editing

\section*{Declaration of generative AI and AI-assisted technologies in the manuscript preparation process}

During the preparation of this work the author(s) used the Gemma~4 (26B-A4B-it) large language model via the llama.cpp inference engine to improve linguistic clarity and reword certain sentences. After using this tool/service, the author(s) reviewed and edited the content as needed and take(s) full responsibility for the content of the published article.

\bibliographystyle{elsarticle-harv}  
\bibliography{references,extrarefs}

\begin{thebibliography}{43}
\expandafter\ifx\csname natexlab\endcsname\relax\def\natexlab#1{#1}\fi
\providecommand{\url}[1]{\texttt{#1}}
\providecommand{\href}[2]{#2}
\providecommand{\path}[1]{#1}
\providecommand{\DOIprefix}{doi:}
\providecommand{\ArXivprefix}{arXiv:}
\providecommand{\URLprefix}{URL: }
\providecommand{\Pubmedprefix}{pmid:}
\providecommand{\doi}[1]{\href{http://dx.doi.org/#1}{\path{#1}}}
\providecommand{\Pubmed}[1]{\href{pmid:#1}{\path{#1}}}
\providecommand{\bibinfo}[2]{#2}
\ifx\xfnm\relax \def\xfnm[#1]{\unskip,\space#1}\fi
\bibitem[{Adams et~al.(2010)Adams, Afonine, Bunkóczi, Chen, Davis, Echols, Headd, Hung, Kapral, Grosse-Kunstleve, McCoy, Moriarty, Oeffner, Read, Richardson, Richardson, Terwilliger and Zwart}]{adams_phenix_2010}
\bibinfo{author}{Adams, P.D.}, \bibinfo{author}{Afonine, P.V.}, \bibinfo{author}{Bunkóczi, G.}, \bibinfo{author}{Chen, V.B.}, \bibinfo{author}{Davis, I.W.}, \bibinfo{author}{Echols, N.}, \bibinfo{author}{Headd, J.J.}, \bibinfo{author}{Hung, L.W.}, \bibinfo{author}{Kapral, G.J.}, \bibinfo{author}{Grosse-Kunstleve, R.W.}, \bibinfo{author}{McCoy, A.J.}, \bibinfo{author}{Moriarty, N.W.}, \bibinfo{author}{Oeffner, R.}, \bibinfo{author}{Read, R.J.}, \bibinfo{author}{Richardson, D.C.}, \bibinfo{author}{Richardson, J.S.}, \bibinfo{author}{Terwilliger, T.C.}, \bibinfo{author}{Zwart, P.H.}, \bibinfo{year}{2010}.
\newblock \bibinfo{title}{\textit{{PHENIX}}: a comprehensive {Python}-based system for macromolecular structure solution}.
\newblock \bibinfo{journal}{Acta Crystallographica Section D Biological Crystallography} \bibinfo{volume}{66}, \bibinfo{pages}{213--221}.
\newblock \URLprefix \url{https://journals.iucr.org/paper?S0907444909052925}, \DOIprefix\doi{10.1107/S0907444909052925}.
\bibitem[{Afonine et~al.(2018)Afonine, Poon, Read, Sobolev, Terwilliger, Urzhumtsev and Adams}]{afonine_real-space_2018}
\bibinfo{author}{Afonine, P.V.}, \bibinfo{author}{Poon, B.K.}, \bibinfo{author}{Read, R.J.}, \bibinfo{author}{Sobolev, O.V.}, \bibinfo{author}{Terwilliger, T.C.}, \bibinfo{author}{Urzhumtsev, A.}, \bibinfo{author}{Adams, P.D.}, \bibinfo{year}{2018}.
\newblock \bibinfo{title}{Real-space refinement in {PHENIX} for cryo-{EM} and crystallography}.
\newblock \bibinfo{journal}{Acta Crystallographica Section D: Structural Biology} \bibinfo{volume}{74}, \bibinfo{pages}{531--544}.
\newblock \URLprefix \url{https://journals.iucr.org/paper?ic5103}, \DOIprefix\doi{10.1107/S2059798318006551}.
\bibitem[{Beckstein et~al.(2018)Beckstein, Seyler and {Kumar, Avishek}}]{beckstein_simulated_2018}
\bibinfo{author}{Beckstein, O.}, \bibinfo{author}{Seyler, S.L.}, \bibinfo{author}{{Kumar, Avishek}}, \bibinfo{year}{2018}.
\newblock \bibinfo{title}{Simulated trajectory ensembles for the closed-to-open transition of adenylate kinase from {DIMS} {MD} and {FRODA}}.
\newblock \URLprefix \url{https://figshare.com/articles/Simulated_trajectory_ensembles_for_the_closed-to-open_transition_of_adenylate_kinase_from_DIMS_MD_and_FRODA/7165306/2}, \DOIprefix\doi{10.6084/M9.FIGSHARE.7165306.V2}.
\bibitem[{Chen and Ludtke(2021)}]{chen_deep_2021}
\bibinfo{author}{Chen, M.}, \bibinfo{author}{Ludtke, S.J.}, \bibinfo{year}{2021}.
\newblock \bibinfo{title}{Deep learning-based mixed-dimensional {Gaussian} mixture model for characterizing variability in cryo-{EM}}.
\newblock \bibinfo{journal}{Nature Methods} \bibinfo{volume}{18}, \bibinfo{pages}{930--936}.
\newblock \URLprefix \url{https://www.nature.com/articles/s41592-021-01220-5}, \DOIprefix\doi{10.1038/s41592-021-01220-5}. \bibinfo{note}{publisher: Nature Publishing Group}.
\bibitem[{Chen et~al.(2024)Chen, Schmid and Chiu}]{chen_improving_2024}
\bibinfo{author}{Chen, M.}, \bibinfo{author}{Schmid, M.F.}, \bibinfo{author}{Chiu, W.}, \bibinfo{year}{2024}.
\newblock \bibinfo{title}{Improving resolution and resolvability of single-particle {cryoEM} structures using {Gaussian} mixture models}.
\newblock \bibinfo{journal}{Nature Methods} \bibinfo{volume}{21}, \bibinfo{pages}{37--40}.
\newblock \URLprefix \url{https://www.nature.com/articles/s41592-023-02082-9}, \DOIprefix\doi{10.1038/s41592-023-02082-9}. \bibinfo{note}{publisher: Nature Publishing Group}.
\bibitem[{Chen et~al.(2023)Chen, Toader and Lederman}]{chen_integrating_2023}
\bibinfo{author}{Chen, M.}, \bibinfo{author}{Toader, B.}, \bibinfo{author}{Lederman, R.}, \bibinfo{year}{2023}.
\newblock \bibinfo{title}{Integrating {Molecular} {Models} {Into} {CryoEM} {Heterogeneity} {Analysis} {Using} {Scalable} {High}-resolution {Deep} {Gaussian} {Mixture} {Models}}.
\newblock \bibinfo{journal}{Journal of Molecular Biology} \bibinfo{volume}{435}, \bibinfo{pages}{168014}.
\newblock \URLprefix \url{https://www.sciencedirect.com/science/article/pii/S0022283623000700}, \DOIprefix\doi{10.1016/j.jmb.2023.168014}.
\bibitem[{Cheng et~al.(2015)Cheng, Grigorieff, Penczek and Walz}]{cheng_primer_2015}
\bibinfo{author}{Cheng, Y.}, \bibinfo{author}{Grigorieff, N.}, \bibinfo{author}{Penczek, P.A.}, \bibinfo{author}{Walz, T.}, \bibinfo{year}{2015}.
\newblock \bibinfo{title}{A {Primer} to {Single}-{Particle} {Cryo}-{Electron} {Microscopy}}.
\newblock \bibinfo{journal}{Cell} \bibinfo{volume}{161}, \bibinfo{pages}{438--449}.
\newblock \URLprefix \url{https://www.cell.com/cell/abstract/S0092-8674(15)00370-0}, \DOIprefix\doi{10.1016/j.cell.2015.03.050}. \bibinfo{note}{publisher: Elsevier}.
\bibitem[{Cowtan(2006)}]{cowtan_buccaneer_2006}
\bibinfo{author}{Cowtan, K.}, \bibinfo{year}{2006}.
\newblock \bibinfo{title}{The \textit{{Buccaneer}} software for automated model building. 1. {Tracing} protein chains}.
\newblock \bibinfo{journal}{Acta Crystallographica Section D Biological Crystallography} \bibinfo{volume}{62}, \bibinfo{pages}{1002--1011}.
\newblock \URLprefix \url{https://journals.iucr.org/paper?S0907444906022116}, \DOIprefix\doi{10.1107/S0907444906022116}.
\bibitem[{Crouse et~al.(2011)Crouse, Willett, Pattipati and Svensson}]{crouse_look_2011}
\bibinfo{author}{Crouse, D.F.}, \bibinfo{author}{Willett, P.}, \bibinfo{author}{Pattipati, K.R.}, \bibinfo{author}{Svensson, L.}, \bibinfo{year}{2011}.
\newblock \bibinfo{title}{A look at {Gaussian} mixture reduction algorithms}.
\newblock \bibinfo{journal}{International Conference on Information Fusion} , \bibinfo{pages}{1--8}.
\bibitem[{Ducrocq et~al.(2024)Ducrocq, Grunewald, Westenhoff and Lindsten}]{ducrocq_cryosphere_2024}
\bibinfo{author}{Ducrocq, G.}, \bibinfo{author}{Grunewald, L.}, \bibinfo{author}{Westenhoff, S.}, \bibinfo{author}{Lindsten, F.}, \bibinfo{year}{2024}.
\newblock \bibinfo{title}{{cryoSPHERE}: {Single}-particle heterogeneous reconstruction from cryo {EM}}.
\newblock \bibinfo{journal}{bioRxiv preprint} \URLprefix \url{https://www.biorxiv.org/content/10.1101/2024.06.19.599686v1}, \DOIprefix\doi{10.1101/2024.06.19.599686}.
\bibitem[{Emsley et~al.(2010)Emsley, Lohkamp, Scott and Cowtan}]{emsley_features_2010}
\bibinfo{author}{Emsley, P.}, \bibinfo{author}{Lohkamp, B.}, \bibinfo{author}{Scott, W.G.}, \bibinfo{author}{Cowtan, K.}, \bibinfo{year}{2010}.
\newblock \bibinfo{title}{Features and development of {Coot}}.
\newblock \bibinfo{journal}{Acta Crystallographica Section D Biological Crystallography} \bibinfo{volume}{66}, \bibinfo{pages}{486--501}.
\newblock \URLprefix \url{https://journals.iucr.org/paper?S0907444910007493}, \DOIprefix\doi{10.1107/S0907444910007493}.
\bibitem[{Gordon et~al.(2012)Gordon, Fedorov, Pruitt and Slipchenko}]{gordon_fragmentation_2012}
\bibinfo{author}{Gordon, M.S.}, \bibinfo{author}{Fedorov, D.G.}, \bibinfo{author}{Pruitt, S.R.}, \bibinfo{author}{Slipchenko, L.V.}, \bibinfo{year}{2012}.
\newblock \bibinfo{title}{Fragmentation {Methods}: {A} {Route} to {Accurate} {Calculations} on {Large} {Systems}}.
\newblock \bibinfo{journal}{Chemical Reviews} \bibinfo{volume}{112}, \bibinfo{pages}{632--672}.
\newblock \URLprefix \url{https://doi.org/10.1021/cr200093j}, \DOIprefix\doi{10.1021/cr200093j}. \bibinfo{note}{publisher: American Chemical Society}.
\bibitem[{Gupta et~al.(2021)Gupta, McCann, Donati and Unser}]{gupta_cryogan_2021}
\bibinfo{author}{Gupta, H.}, \bibinfo{author}{McCann, M.T.}, \bibinfo{author}{Donati, L.}, \bibinfo{author}{Unser, M.}, \bibinfo{year}{2021}.
\newblock \bibinfo{title}{{CryoGAN}: {A} {New} {Reconstruction} {Paradigm} for {Single}-{Particle} {Cryo}-{EM} {Via} {Deep} {Adversarial} {Learning}}.
\newblock \bibinfo{journal}{IEEE Transactions on Computational Imaging} \bibinfo{volume}{7}, \bibinfo{pages}{759--774}.
\newblock \URLprefix \url{https://ieeexplore.ieee.org/document/9483649}, \DOIprefix\doi{10.1109/TCI.2021.3096491}.
\bibitem[{Gupta et~al.(2020)Gupta, Phan, Yoo and Unser}]{gupta_multi-cryogan_2020}
\bibinfo{author}{Gupta, H.}, \bibinfo{author}{Phan, T.H.}, \bibinfo{author}{Yoo, J.}, \bibinfo{author}{Unser, M.}, \bibinfo{year}{2020}.
\newblock \bibinfo{title}{Multi-{CryoGAN}: {Reconstruction} of {Continuous} {Conformations} in {Cryo}-{EM} {Using} {Generative} {Adversarial} {Networks}}, in: \bibinfo{booktitle}{Computer {Vision} -- {ECCV} 2020 {Workshops}}, \bibinfo{publisher}{Springer International Publishing}. pp. \bibinfo{pages}{429--444}.
\newblock \URLprefix \url{https://openreview.net/forum?id=5PSL-CjHeP4}.
\bibitem[{Hu et~al.(2011)Hu, Lundgren and Niemi}]{hu_discrete_2011}
\bibinfo{author}{Hu, S.}, \bibinfo{author}{Lundgren, M.}, \bibinfo{author}{Niemi, A.J.}, \bibinfo{year}{2011}.
\newblock \bibinfo{title}{Discrete {Frenet} frame, inflection point solitons, and curve visualization with applications to folded proteins}.
\newblock \bibinfo{journal}{Physical Review E} \bibinfo{volume}{83}, \bibinfo{pages}{061908}.
\newblock \URLprefix \url{https://link.aps.org/doi/10.1103/PhysRevE.83.061908}, \DOIprefix\doi{10.1103/PhysRevE.83.061908}. \bibinfo{note}{publisher: American Physical Society}.
\bibitem[{Jamali et~al.(2023)Jamali, Kimanius and Scheres}]{jamali_graph_2023}
\bibinfo{author}{Jamali, K.}, \bibinfo{author}{Kimanius, D.}, \bibinfo{author}{Scheres, S.H.}, \bibinfo{year}{2023}.
\newblock \bibinfo{title}{A {Graph} {Neural} {Network} {Approach} to {Automated} {Model} {Building} in {Cryo}-{EM} {Maps}}.
\newblock \URLprefix \url{https://openreview.net/forum?id=65XDF_nwI61}.
\bibitem[{Jamali et~al.(2024)Jamali, Käll, Zhang, Brown, Kimanius and Scheres}]{jamali_automated_2024}
\bibinfo{author}{Jamali, K.}, \bibinfo{author}{Käll, L.}, \bibinfo{author}{Zhang, R.}, \bibinfo{author}{Brown, A.}, \bibinfo{author}{Kimanius, D.}, \bibinfo{author}{Scheres, S.H.W.}, \bibinfo{year}{2024}.
\newblock \bibinfo{title}{Automated model building and protein identification in cryo-{EM} maps}.
\newblock \bibinfo{journal}{Nature} \bibinfo{volume}{628}, \bibinfo{pages}{450--457}.
\newblock \URLprefix \url{https://www.nature.com/articles/s41586-024-07215-4}, \DOIprefix\doi{10.1038/s41586-024-07215-4}. \bibinfo{note}{publisher: Nature Publishing Group}.
\bibitem[{Kirkland(1998)}]{kirkland_advanced_1998}
\bibinfo{author}{Kirkland, E.J.}, \bibinfo{year}{1998}.
\newblock \bibinfo{title}{Advanced computing in electron microscopy}.
\newblock \bibinfo{publisher}{Plenum Press}, \bibinfo{address}{New York and London}.
\bibitem[{Kmiecik et~al.(2016)Kmiecik, Gront, Kolinski, Wieteska, Dawid and Kolinski}]{kmiecik_coarse-grained_2016}
\bibinfo{author}{Kmiecik, S.}, \bibinfo{author}{Gront, D.}, \bibinfo{author}{Kolinski, M.}, \bibinfo{author}{Wieteska, L.}, \bibinfo{author}{Dawid, A.E.}, \bibinfo{author}{Kolinski, A.}, \bibinfo{year}{2016}.
\newblock \bibinfo{title}{Coarse-{Grained} {Protein} {Models} and {Their} {Applications}}.
\newblock \bibinfo{journal}{Chemical Reviews} \bibinfo{volume}{116}, \bibinfo{pages}{7898--7936}.
\newblock \URLprefix \url{https://pubs.acs.org/doi/10.1021/acs.chemrev.6b00163}, \DOIprefix\doi{10.1021/acs.chemrev.6b00163}.
\bibitem[{Kufareva and Abagyan(2012)}]{kufareva_methods_2012}
\bibinfo{author}{Kufareva, I.}, \bibinfo{author}{Abagyan, R.}, \bibinfo{year}{2012}.
\newblock \bibinfo{title}{Methods of protein structure comparison}.
\newblock \bibinfo{journal}{Methods in Molecular Biology (Clifton, N.J.)} \bibinfo{volume}{857}, \bibinfo{pages}{231--257}.
\newblock \DOIprefix\doi{10.1007/978-1-61779-588-6_10}.
\bibitem[{Levy et~al.(2022)Levy, Poitevin, Martel, Nashed, Peck, Miolane, Ratner, Dunne and Wetzstein}]{levy_cryoai_2022}
\bibinfo{author}{Levy, A.}, \bibinfo{author}{Poitevin, F.}, \bibinfo{author}{Martel, J.}, \bibinfo{author}{Nashed, Y.}, \bibinfo{author}{Peck, A.}, \bibinfo{author}{Miolane, N.}, \bibinfo{author}{Ratner, D.}, \bibinfo{author}{Dunne, M.}, \bibinfo{author}{Wetzstein, G.}, \bibinfo{year}{2022}.
\newblock \bibinfo{title}{{CryoAI}: {Amortized} {Inference} of {Poses} for {Ab} {Initio} {Reconstruction} of {3D} {Molecular} {Volumes} from {Real} {Cryo}-{EM} {Images}}, in: \bibinfo{booktitle}{Computer {Vision} – {ECCV} 2022: 17th {European} {Conference}, {Tel} {Aviv}, {Israel}, {October} 23–27, 2022, {Proceedings}, {Part} {XXI}}, \bibinfo{publisher}{Springer-Verlag}, \bibinfo{address}{Berlin, Heidelberg}. pp. \bibinfo{pages}{540--557}.
\newblock \URLprefix \url{https://doi.org/10.1007/978-3-031-19803-8_32}, \DOIprefix\doi{10.1007/978-3-031-19803-8_32}.
\bibitem[{Li et~al.(2018)Li, Xu, Taylor, Studer and Goldstein}]{li_visualizing_2018}
\bibinfo{author}{Li, H.}, \bibinfo{author}{Xu, Z.}, \bibinfo{author}{Taylor, G.}, \bibinfo{author}{Studer, C.}, \bibinfo{author}{Goldstein, T.}, \bibinfo{year}{2018}.
\newblock \bibinfo{title}{Visualizing the {Loss} {Landscape} of {Neural} {Nets}}.
\newblock \URLprefix \url{http://arxiv.org/abs/1712.09913}. \bibinfo{note}{arXiv:1712.09913 [cs, stat]}.
\bibitem[{Li et~al.(2024)Li, Zhou, Yuan, Ye and Gu}]{li_cryostar_2024}
\bibinfo{author}{Li, Y.}, \bibinfo{author}{Zhou, Y.}, \bibinfo{author}{Yuan, J.}, \bibinfo{author}{Ye, F.}, \bibinfo{author}{Gu, Q.}, \bibinfo{year}{2024}.
\newblock \bibinfo{title}{{CryoSTAR}: leveraging structural priors and constraints for cryo-{EM} heterogeneous reconstruction}.
\newblock \bibinfo{journal}{Nature Methods} \bibinfo{volume}{21}, \bibinfo{pages}{2318--2326}.
\newblock \URLprefix \url{https://www.nature.com/articles/s41592-024-02486-1}, \DOIprefix\doi{10.1038/s41592-024-02486-1}.
\bibitem[{Liebschner et~al.(2019)Liebschner, Afonine, Baker, Bunkóczi, Chen, Croll, Hintze, Hung, Jain, McCoy, Moriarty, Oeffner, Poon, Prisant, Read, Richardson, Richardson, Sammito, Sobolev, Stockwell, Terwilliger, Urzhumtsev, Videau, Williams and Adams}]{liebschner_macromolecular_2019}
\bibinfo{author}{Liebschner, D.}, \bibinfo{author}{Afonine, P.V.}, \bibinfo{author}{Baker, M.L.}, \bibinfo{author}{Bunkóczi, G.}, \bibinfo{author}{Chen, V.B.}, \bibinfo{author}{Croll, T.I.}, \bibinfo{author}{Hintze, B.}, \bibinfo{author}{Hung, L.W.}, \bibinfo{author}{Jain, S.}, \bibinfo{author}{McCoy, A.J.}, \bibinfo{author}{Moriarty, N.W.}, \bibinfo{author}{Oeffner, R.D.}, \bibinfo{author}{Poon, B.K.}, \bibinfo{author}{Prisant, M.G.}, \bibinfo{author}{Read, R.J.}, \bibinfo{author}{Richardson, J.S.}, \bibinfo{author}{Richardson, D.C.}, \bibinfo{author}{Sammito, M.D.}, \bibinfo{author}{Sobolev, O.V.}, \bibinfo{author}{Stockwell, D.H.}, \bibinfo{author}{Terwilliger, T.C.}, \bibinfo{author}{Urzhumtsev, A.G.}, \bibinfo{author}{Videau, L.L.}, \bibinfo{author}{Williams, C.J.}, \bibinfo{author}{Adams, P.D.}, \bibinfo{year}{2019}.
\newblock \bibinfo{title}{Macromolecular structure determination using {X}-rays, neutrons and electrons: recent developments in {Phenix}}.
\newblock \bibinfo{journal}{Acta Crystallographica. Section D, Structural Biology} \bibinfo{volume}{75}, \bibinfo{pages}{861--877}.
\newblock \URLprefix \url{https://pmc.ncbi.nlm.nih.gov/articles/PMC6778852/}, \DOIprefix\doi{10.1107/S2059798319011471}.
\bibitem[{Lobato and Van~Dyck(2015)}]{lobato_multem_2015}
\bibinfo{author}{Lobato, I.}, \bibinfo{author}{Van~Dyck, D.}, \bibinfo{year}{2015}.
\newblock \bibinfo{title}{{MULTEM}: {A} new multislice program to perform accurate and fast electron diffraction and imaging simulations using {Graphics} {Processing} {Units} with {CUDA}}.
\newblock \bibinfo{journal}{Ultramicroscopy} \bibinfo{volume}{156}, \bibinfo{pages}{9--17}.
\newblock \URLprefix \url{https://www.sciencedirect.com/science/article/pii/S030439911500100X}, \DOIprefix\doi{10.1016/j.ultramic.2015.04.016}.
\bibitem[{Milne et~al.(2013)Milne, Borgnia, Bartesaghi, Tran, Earl, Schauder, Lengyel, Pierson, Patwardhan and Subramaniam}]{milne_cryo-electron_2013}
\bibinfo{author}{Milne, J.L.S.}, \bibinfo{author}{Borgnia, M.J.}, \bibinfo{author}{Bartesaghi, A.}, \bibinfo{author}{Tran, E.E.H.}, \bibinfo{author}{Earl, L.A.}, \bibinfo{author}{Schauder, D.M.}, \bibinfo{author}{Lengyel, J.}, \bibinfo{author}{Pierson, J.}, \bibinfo{author}{Patwardhan, A.}, \bibinfo{author}{Subramaniam, S.}, \bibinfo{year}{2013}.
\newblock \bibinfo{title}{Cryo-electron microscopy – a primer for the non-microscopist}.
\newblock \bibinfo{journal}{The FEBS Journal} \bibinfo{volume}{280}, \bibinfo{pages}{28--45}.
\newblock \URLprefix \url{https://onlinelibrary.wiley.com/doi/abs/10.1111/febs.12078}, \DOIprefix\doi{10.1111/febs.12078}.
\bibitem[{Nashed et~al.(2021)Nashed, Poitevin, Gupta, Woollard, Kagan, Yoon and Ratner}]{nashed_cryoposenet_2021}
\bibinfo{author}{Nashed, Y.S.G.}, \bibinfo{author}{Poitevin, F.}, \bibinfo{author}{Gupta, H.}, \bibinfo{author}{Woollard, G.}, \bibinfo{author}{Kagan, M.}, \bibinfo{author}{Yoon, C.H.}, \bibinfo{author}{Ratner, D.}, \bibinfo{year}{2021}.
\newblock \bibinfo{title}{{CryoPoseNet}: {End}-to-{End} {Simultaneous} {Learning} of {Single}-particle {Orientation} and {3D} {Map} {Reconstruction} from {Cryo}-electron {Microscopy} {Data}}, in: \bibinfo{booktitle}{2021 {IEEE}/{CVF} {International} {Conference} on {Computer} {Vision} {Workshops} ({ICCVW})}, \bibinfo{publisher}{IEEE}, \bibinfo{address}{Montreal, BC, Canada}. pp. \bibinfo{pages}{4049--4059}.
\newblock \URLprefix \url{https://ieeexplore.ieee.org/document/9607470/}, \DOIprefix\doi{10.1109/ICCVW54120.2021.00452}.
\bibitem[{Parkhurst et~al.(2021)Parkhurst, Dumoux, Basham, Clare, Siebert, Varslot, Kirkland, Naismith and Evans}]{parkhurst_parakeet_2021}
\bibinfo{author}{Parkhurst, J.M.}, \bibinfo{author}{Dumoux, M.}, \bibinfo{author}{Basham, M.}, \bibinfo{author}{Clare, D.}, \bibinfo{author}{Siebert, C.A.}, \bibinfo{author}{Varslot, T.}, \bibinfo{author}{Kirkland, A.}, \bibinfo{author}{Naismith, J.H.}, \bibinfo{author}{Evans, G.}, \bibinfo{year}{2021}.
\newblock \bibinfo{title}{Parakeet: a digital twin software pipeline to assess the impact of experimental parameters on tomographic reconstructions for cryo-electron tomography}.
\newblock \bibinfo{journal}{Open Biology} \bibinfo{volume}{11}, \bibinfo{pages}{210160}.
\newblock \URLprefix \url{https://royalsocietypublishing.org/doi/10.1098/rsob.210160}, \DOIprefix\doi{10.1098/rsob.210160}. \bibinfo{note}{publisher: Royal Society}.
\bibitem[{Paszke et~al.(2019)Paszke, Gross, Massa, Lerer, Bradbury, Chanan, Killeen, Lin, Gimelshein, Antiga, Desmaison, Köpf, Yang, DeVito, Raison, Tejani, Chilamkurthy, Steiner, Fang, Bai and Chintala}]{paszke_pytorch_2019}
\bibinfo{author}{Paszke, A.}, \bibinfo{author}{Gross, S.}, \bibinfo{author}{Massa, F.}, \bibinfo{author}{Lerer, A.}, \bibinfo{author}{Bradbury, J.}, \bibinfo{author}{Chanan, G.}, \bibinfo{author}{Killeen, T.}, \bibinfo{author}{Lin, Z.}, \bibinfo{author}{Gimelshein, N.}, \bibinfo{author}{Antiga, L.}, \bibinfo{author}{Desmaison, A.}, \bibinfo{author}{Köpf, A.}, \bibinfo{author}{Yang, E.}, \bibinfo{author}{DeVito, Z.}, \bibinfo{author}{Raison, M.}, \bibinfo{author}{Tejani, A.}, \bibinfo{author}{Chilamkurthy, S.}, \bibinfo{author}{Steiner, B.}, \bibinfo{author}{Fang, L.}, \bibinfo{author}{Bai, J.}, \bibinfo{author}{Chintala, S.}, \bibinfo{year}{2019}.
\newblock \bibinfo{title}{{PyTorch}: an imperative style, high-performance deep learning library}, in: \bibinfo{booktitle}{Proceedings of the 33rd {International} {Conference} on {Neural} {Information} {Processing} {Systems}}. \bibinfo{publisher}{Curran Associates Inc.}, \bibinfo{address}{Red Hook, NY, USA}, pp. \bibinfo{pages}{8024--8035}.
\bibitem[{Punjani and Fleet(2023)}]{punjani_3dflex_2023}
\bibinfo{author}{Punjani, A.}, \bibinfo{author}{Fleet, D.J.}, \bibinfo{year}{2023}.
\newblock \bibinfo{title}{{3DFlex}: determining structure and motion of flexible proteins from cryo-{EM}}.
\newblock \bibinfo{journal}{Nature Methods} \bibinfo{volume}{20}, \bibinfo{pages}{860--870}.
\newblock \URLprefix \url{https://www.nature.com/articles/s41592-023-01853-8}, \DOIprefix\doi{10.1038/s41592-023-01853-8}. \bibinfo{note}{publisher: Nature Publishing Group}.
\bibitem[{Punjani et~al.(2017)Punjani, Rubinstein, Fleet and Brubaker}]{punjani_cryosparc_2017}
\bibinfo{author}{Punjani, A.}, \bibinfo{author}{Rubinstein, J.L.}, \bibinfo{author}{Fleet, D.J.}, \bibinfo{author}{Brubaker, M.A.}, \bibinfo{year}{2017}.
\newblock \bibinfo{title}{{cryoSPARC}: algorithms for rapid unsupervised cryo-{EM} structure determination}.
\newblock \bibinfo{journal}{Nature Methods} \bibinfo{volume}{14}, \bibinfo{pages}{290--296}.
\newblock \URLprefix \url{https://www.nature.com/articles/nmeth.4169}, \DOIprefix\doi{10.1038/nmeth.4169}. \bibinfo{note}{publisher: Nature Publishing Group}.
\bibitem[{Rosenbaum et~al.(2021)Rosenbaum, Garnelo, Zielinski, Beattie, Clancy, Huber, Kohli, Senior, Jumper, Doersch, Eslami, Ronneberger and Adler}]{rosenbaum_inferring_2021}
\bibinfo{author}{Rosenbaum, D.}, \bibinfo{author}{Garnelo, M.}, \bibinfo{author}{Zielinski, M.}, \bibinfo{author}{Beattie, C.}, \bibinfo{author}{Clancy, E.}, \bibinfo{author}{Huber, A.}, \bibinfo{author}{Kohli, P.}, \bibinfo{author}{Senior, A.}, \bibinfo{author}{Jumper, J.}, \bibinfo{author}{Doersch, C.}, \bibinfo{author}{Eslami, S.}, \bibinfo{author}{Ronneberger, O.}, \bibinfo{author}{Adler, J.}, \bibinfo{year}{2021}.
\newblock \bibinfo{title}{Inferring a {Continuous} {Distribution} of {Atom} {Coordinates} from {Cryo}-{EM} {Images} using {VAEs}}.
\newblock \bibinfo{journal}{arXiv preprint} \URLprefix \url{https://arxiv.org/abs/2106.14108}.
\bibitem[{Scheres(2012)}]{scheres_relion_2012}
\bibinfo{author}{Scheres, S.H.}, \bibinfo{year}{2012}.
\newblock \bibinfo{title}{{RELION}: {Implementation} of a {Bayesian} approach to cryo-{EM} structure determination}.
\newblock \bibinfo{journal}{Journal of Structural Biology} \bibinfo{volume}{180}, \bibinfo{pages}{519--530}.
\newblock \URLprefix \url{https://www.ncbi.nlm.nih.gov/pmc/articles/PMC3690530/}, \DOIprefix\doi{10.1016/j.jsb.2012.09.006}.
\bibitem[{Schwab et~al.(2024)Schwab, Kimanius, Burt, Dendooven and Scheres}]{schwab_dynamight_2024}
\bibinfo{author}{Schwab, J.}, \bibinfo{author}{Kimanius, D.}, \bibinfo{author}{Burt, A.}, \bibinfo{author}{Dendooven, T.}, \bibinfo{author}{Scheres, S.H.W.}, \bibinfo{year}{2024}.
\newblock \bibinfo{title}{{DynaMight}: estimating molecular motions with improved reconstruction from cryo-{EM} images}.
\newblock \bibinfo{journal}{Nature Methods} , \bibinfo{pages}{1--8}\URLprefix \url{https://www.nature.com/articles/s41592-024-02377-5}, \DOIprefix\doi{10.1038/s41592-024-02377-5}. \bibinfo{note}{publisher: Nature Publishing Group}.
\bibitem[{Sigworth(2022)}]{sigworth_aemcoderepository_2022}
\bibinfo{author}{Sigworth, F.}, \bibinfo{year}{2022}.
\newblock \bibinfo{title}{{aEMCodeRepository}}.
\newblock \bibinfo{note}{Published: Software Heritage archive}.
\bibitem[{Terashi et~al.(2024)Terashi, Wang, Prasad, Nakamura and Kihara}]{terashi_deepmainmast_2024}
\bibinfo{author}{Terashi, G.}, \bibinfo{author}{Wang, X.}, \bibinfo{author}{Prasad, D.}, \bibinfo{author}{Nakamura, T.}, \bibinfo{author}{Kihara, D.}, \bibinfo{year}{2024}.
\newblock \bibinfo{title}{{DeepMainmast}: integrated protocol of protein structure modeling for cryo-{EM} with deep learning and structure prediction}.
\newblock \bibinfo{journal}{Nature Methods} \bibinfo{volume}{21}, \bibinfo{pages}{122--131}.
\newblock \URLprefix \url{https://www.nature.com/articles/s41592-023-02099-0}, \DOIprefix\doi{10.1038/s41592-023-02099-0}. \bibinfo{note}{publisher: Nature Publishing Group}.
\bibitem[{Terwilliger et~al.(2008)Terwilliger, Grosse-Kunstleve, Afonine, Moriarty, Zwart, Hung, Read and Adams}]{terwilliger_iterative_2008}
\bibinfo{author}{Terwilliger, T.C.}, \bibinfo{author}{Grosse-Kunstleve, R.W.}, \bibinfo{author}{Afonine, P.V.}, \bibinfo{author}{Moriarty, N.W.}, \bibinfo{author}{Zwart, P.H.}, \bibinfo{author}{Hung, L.W.}, \bibinfo{author}{Read, R.J.}, \bibinfo{author}{Adams, P.D.}, \bibinfo{year}{2008}.
\newblock \bibinfo{title}{Iterative model building, structure refinement and density modification with the {PHENIX} {AutoBuild} wizard}.
\newblock \bibinfo{journal}{Acta Crystallographica. Section D, Biological Crystallography} \bibinfo{volume}{64}, \bibinfo{pages}{61--69}.
\newblock \DOIprefix\doi{10.1107/S090744490705024X}.
\bibitem[{Vander Meersche et~al.(2024)Vander Meersche, Cretin, Gheeraert, Gelly and Galochkina}]{vandermeersche_atlas_2024}
\bibinfo{author}{Vander Meersche, Y.}, \bibinfo{author}{Cretin, G.}, \bibinfo{author}{Gheeraert, A.}, \bibinfo{author}{Gelly, J.C.}, \bibinfo{author}{Galochkina, T.}, \bibinfo{year}{2024}.
\newblock \bibinfo{title}{{ATLAS}: protein flexibility description from atomistic molecular dynamics simulations}.
\newblock \bibinfo{journal}{Nucleic Acids Research} \bibinfo{volume}{52}, \bibinfo{pages}{D384--D392}.
\newblock \URLprefix \url{https://doi.org/10.1093/nar/gkad1084}, \DOIprefix\doi{10.1093/nar/gkad1084}.
\bibitem[{Zhang et~al.(2022)Zhang, Zhang, Freddolino and Zhang}]{zhang_cr-i-tasser_2022}
\bibinfo{author}{Zhang, X.}, \bibinfo{author}{Zhang, B.}, \bibinfo{author}{Freddolino, P.L.}, \bibinfo{author}{Zhang, Y.}, \bibinfo{year}{2022}.
\newblock \bibinfo{title}{{CR}-{I}-{TASSER}: assemble protein structures from cryo-{EM} density maps using deep convolutional neural networks}.
\newblock \bibinfo{journal}{Nature Methods} \bibinfo{volume}{19}, \bibinfo{pages}{195--204}.
\newblock \URLprefix \url{https://www.nature.com/articles/s41592-021-01389-9}, \DOIprefix\doi{10.1038/s41592-021-01389-9}. \bibinfo{note}{publisher: Nature Publishing Group}.
\bibitem[{Zhang and Skolnick(2004)}]{zhang_scoring_2004}
\bibinfo{author}{Zhang, Y.}, \bibinfo{author}{Skolnick, J.}, \bibinfo{year}{2004}.
\newblock \bibinfo{title}{Scoring function for automated assessment of protein structure template quality}.
\newblock \bibinfo{journal}{Proteins: Structure, Function, and Bioinformatics} \bibinfo{volume}{57}, \bibinfo{pages}{702--710}.
\newblock \URLprefix \url{https://onlinelibrary.wiley.com/doi/10.1002/prot.20264}, \DOIprefix\doi{10.1002/prot.20264}.
\bibitem[{Zhong et~al.(2021a)Zhong, Bepler, Berger and Davis}]{zhong_cryodrgn_2021}
\bibinfo{author}{Zhong, E.}, \bibinfo{author}{Bepler, T.}, \bibinfo{author}{Berger, B.}, \bibinfo{author}{Davis, J.H.}, \bibinfo{year}{2021}a.
\newblock \bibinfo{title}{{CryoDRGN}: reconstruction of heterogeneous cryo-{EM} structures using neural networks}.
\newblock \bibinfo{journal}{Nature Methods} \bibinfo{volume}{18}, \bibinfo{pages}{176--185}.
\newblock \URLprefix \url{https://www.nature.com/articles/s41592-020-01049-4}, \DOIprefix\doi{10.1038/s41592-020-01049-4}. \bibinfo{note}{publisher: Nature Publishing Group}.
\bibitem[{Zhong et~al.(2021b)Zhong, Lerer, Davis and Berger}]{zhong_exploring_2021}
\bibinfo{author}{Zhong, E.}, \bibinfo{author}{Lerer, A.}, \bibinfo{author}{Davis, J.}, \bibinfo{author}{Berger, B.}, \bibinfo{year}{2021}b.
\newblock \bibinfo{title}{Exploring generative atomic models in cryo-{EM} reconstruction}, in: \bibinfo{booktitle}{{NeurIPS} {Workshop} on {Machine} {Learning} for {Structural} {Biology} ({MLSB})}.
\newblock \URLprefix \url{https://www.semanticscholar.org/paper/Exploring-generative-atomic-models-in-cryo-EM-Zhong-Lerer/0bafec3e3c1918cac8deb0f47f1c9defa944d75b}.
\bibitem[{Zhong et~al.(2021c)Zhong, Lerer, Davis and Berger}]{zhong_cryodrgn2_2021}
\bibinfo{author}{Zhong, E.}, \bibinfo{author}{Lerer, A.}, \bibinfo{author}{Davis, J.H.}, \bibinfo{author}{Berger, B.}, \bibinfo{year}{2021}c.
\newblock \bibinfo{title}{{CryoDRGN2}: {Ab} initio neural reconstruction of {3D} protein structures from real cryo-{EM} images}, in: \bibinfo{booktitle}{2021 {IEEE}/{CVF} {International} {Conference} on {Computer} {Vision} ({ICCV})}, pp. \bibinfo{pages}{4046--4055}.
\newblock \URLprefix \url{https://ieeexplore.ieee.org/document/9710804/?arnumber=9710804}, \DOIprefix\doi{10.1109/ICCV48922.2021.00403}. \bibinfo{note}{iSSN: 2380-7504}.

\end{thebibliography}

\appendix
\section{Expression of the Backbone Chain through Frenet Frames}\label{app:manif}

Recall that the atomic coordinates of the backbone of a single protein chain can be denoted by $B$ as in \eqref{eq:SeqOfTriplets}.
Each atom in the backbone chain is connected to the next atom in the backbone chain, and thus its position can be defined in terms of the previous atom in the backbone chain.  This is represented by:

\begin{align}
\mathbf{r}_{i+1} = \mathbf{r}_{i} + \delta_{i} \mathbf{t}_{i} \label{eq:coordsfromfrenet}
\end{align}
where $\delta_0,\delta_1,\ldots,\delta_{3n-2}\in \mathbb{R}$ denotes bond lengths and the $\mathbf{t}_{0},\mathbf{t}_{1},\ldots,\mathbf{t}_{3n-2}\in S^2$ represents bond directions.

We can parametrise the bond orientations by pairs of torsion angles and bond angles through the use of the discrete Frenet frame \citep{hu_discrete_2011}:
\begin{alignat*}{2}
    \delta_{i} &=\left\Vert \mathbf{r}_{i+1}-\mathbf{r}_{i}\right\Vert
    &\qquad
    \mathbf{b}_{i} &=\frac{\mathbf{t}_{i-1}\times\mathbf{t}_{i}}{\left\Vert \mathbf{t}_{i-1}\times\mathbf{t}_{i}\right\Vert}
    \\
    \mathbf{t}_i &= \frac{\mathbf{r}_{i+1} - \mathbf{r}_i}{\delta_{i}} 
    &\qquad
    \mathbf{n}_{i} &=\mathbf{b}_{i}\times\mathbf{t}_{i},
\end{alignat*}
where $\times$ denotes the vector cross product. This is a discrete version of the Frenet--Serret frame, which is used to describe a curve in 3D space, as well as how it torsions. The vector $\mathbf{t}_i$ is known as the tangent which points in the direction of motion. The binormal and normal vectors $\mathbf{b}_{i}$ and $\mathbf{n}_{i}$ are mutually orthogonal with $\mathbf{t}_i$  and describe the torsion of the curve. Intuitively, at each vertex $i$ of the discrete curve, $\mathbf{t}_{i-1}$ and $\mathbf{t}_i$ are the directions of entry and exit into that vertex, $\mathbf{b}_{i}$ is the axis of rotation that maps $\mathbf{t}_{i-1}$ to $\mathbf{t}_i$,  and $\mathbf{n}_{i}$ points in the direction of bending.   Taken together, $\mathbf{t}_i$, $\mathbf{b}_{i}$ and $\mathbf{n}_{i}$ form a local frame at each bend in the curve -- a local coordinate system. We made sure that the Frenet frame was uniquely defined by setting a fixed convention for the start/end vectors $\mathbf{b}_{1},\mathbf{b}_{N}$, $\mathbf{n}_{1},\mathbf{n}_{N}$, and fixing the order of traversal of the chain.

Each triplet $F_{i}=(\mathbf{n}_{i},\mathbf{b}_{i},\mathbf{t}_{i})^{\top}$ is related to the previous triplet via the relationship $F_{i} = A_{i} F_{i-1}$, where:
\begin{equation}
A_{i}	= \begin{bmatrix}\cos\kappa_{i}\cos\tau_{i} & \cos\kappa_{i}\sin\tau_{i} & -\sin\kappa_{i}\\
-\sin\tau_{i} & \cos\tau_{i} & 0\\
\sin\kappa_{i}\cos\tau_{i} & \sin\kappa_{i}\sin\tau_{i} & \cos\kappa_{i}
\end{bmatrix} \label{eq:recursion-rotate}
\end{equation}
for bond angle $\kappa_{i}$ and torsion angle $\tau_{i}$, both relating to the $i$th bond, where:
\begin{alignat*}{2}
    \kappa_i &= \arccos(\mathbf{t}_i \cdot \mathbf{t}_{i-1}) \in [0, \pi] & \quad \tau_1 &= 0 \\
    \tau_i &= \operatorname{atan2}\left( (\mathbf{b}_{i-1} \times \mathbf{b}_i) \cdot \mathbf{t}_i, \mathbf{b}_i \cdot \mathbf{b}_{i-1} \right) \in (-\pi, \pi] & \quad\text{for } i &> 1
\end{alignat*}
See Figure~\ref{fig:rot-torsions} for an illustration. The cross product relationship $\mathbf{n}_{i} =\mathbf{b}_{i}\times\mathbf{t}_{i}$ is preserved because $A_{i}\in \text{SO}(3)$ and the cross product is preserved under SO(3).

The atomic coordinates $\mathbf{r}_0, \mathbf{r}_1, \ldots$ can be obtained back from the Frenet frames by using $F_{i} = A_{i} F_{i-1}$, the matrix in Eq.~\eqref{eq:recursion-rotate}, and the recursion relation in Eq.~\eqref{eq:coordsfromfrenet}.
Thus the state of the chain can be fully described by: the bond lengths $\delta_i$, the position of the first atom in the chain $\mathbf{r}_0$, $F_0=(\mathbf{n}_{0},\mathbf{b}_{0},\mathbf{t}_{0})^{\top}$, the torsions $\tau = [\tau_{1},\tau_{2},\tau_{3},\ldots,\tau_{3n-1}]$ and bond angles $\kappa = [\kappa_{1},\kappa_{2},\ldots,\kappa_{3n-1}]$ (the action of rotation along the pairs of bond angles can be parametrised as elements of $S^2$). Since the bond lengths $\delta_i$ are assumed to be fixed, \eqref{eq:recursion-rotate} and the recursion $F_{i} = A_{i} F_{i-1}$ relates the bond angles back to the sequence $\{{t}_\mathbf{i}\}_{i=0}^{3n-2}$, which then can be converted back to the Cartesian atomic coordinates $B_0$ by \eqref{eq:coordsfromfrenet}.

Observe that changes in orientation of the first bond $F_0$ and position $\mathbf{r}_0$ of the first atom only result in rotation and rigid body translations of the molecule, not any change in its internal conformation. Thus, the conformation of the molecule is fully described by the torsion and bond angles. 
This simplifies the expression of deformations of the protein chain through changes in bond torsion and angles: since the bond lengths are assumed to be constant, all deformations are changes in torsions and bond angles, $\tau_i$, $\kappa_i$, starting point $\mathbf{r}_0$ and first Frenet frame $F_0$. 

For a single protein chain of $n$ residues, we thus define the\emph{ manifold of protein chains} $\mathcal{M}_n\subseteq \mathbb{R}^3 \times \mathrm{SO}(3) \times \mathbb{R}^{6n - 4}$, where a point on $\mathcal{M}_n$ is given by the tuple $\bigl( \mathbf{r}_0, F_0, \{ \kappa_i \}_{i=1}^{3n-2}, \{ \tau_i \}_{i=1}^{3n-2} \bigr)$.

\section{Mathematical formulation of Gaussian Dictionary Element Position in Relation to Backbone Conformation \label{app:relgauss}}

This section gives the full mathematical formulation of how the position and orientation of each Gaussian changes in relation to the coordinates of the three consecutive backbone atoms. We thus define a local reference frame (below) at each triplet of backbone atoms, based on the positions of the $N-C_\alpha$ and $C_\alpha - C$ bonds in the template conformation. This local reference frame is an orthonormal basis derived from the orientations of these bonds. 

The frame or coordinate system is attached to the positions of the $N,C_\alpha$ and $C$ atoms in the backbone triplet and changes as the position of the atoms in the triplet change -- the orthonormal basis vectors rotate as the bond orientations change -- but the orientation and the position of the Gaussian remains constant within the local reference frame.

More formally, let $\mathbf{r}_{3i}$, $\mathbf{r}_{3i+1}$, and $\mathbf{r}_{3i+2}$ be position vectors of the $N$, $C_\alpha$ and $C$ atoms defining a residue.
Now define
\begin{align*}
\mathbf{u} &= \mathbf{r}_{3i} - \mathbf{r}_{3i+1} \quad &\text{(vector pointing from $C_\alpha$ to $N$)} \\
\mathbf{v} &= \mathbf{r}_{3i+2} - \mathbf{r}_{3i+1} \quad &\text{(vector pointing from $C_\alpha$ to $C$)}\\  
\end{align*}
From this, we create an orthonormal basis using a modified Gram--Schmidt orthonomalisation process:
\begin{alignat*}{3}
\mathbf{u}' &= \mathbf{u} + \mathbf{v}
&\qquad\qquad
\mathbf{e}_0 &= \frac{\mathbf{u}'}{||\mathbf{u}'||}
\\
\mathbf{v}' &=  \mathbf{v} - \mathrm{proj}_{\mathbf{u}'} \mathbf{v}
&\qquad\qquad
\mathbf{e}_1 &= \frac{\mathbf{v}'}{||\mathbf{v}'||}
\\
\mathbf{w} &= \mathbf{u}' \times \mathbf{v}'
&
\mathbf{e}_2 &= \frac{\mathbf{w}}{||\mathbf{w}||}
\end{alignat*}
The vector $\mathbf{u}' = \mathbf{u} + \mathbf{v}$ is used to increase stability, in case the angle between $\mathbf{u}$ and $\mathbf{v}$ becomes greater than $\pi$. This yields an orthonormal basis $\mathbf{e}_{0},\mathbf{e}_{1}, \mathbf{e}_{2}$, which form a coordinate system. The origin of this coordinate system located at $\mathbf{r}_{3i+1}$, the position of the $C_\alpha$ atom.  The origin and the orthonormal basis $(\mathbf{r}_{3i+1},(\mathbf{e}_{0},\mathbf{e}_{1}, \mathbf{e}_{2}))$ shall be referred to as the \emph{local frame} of residue $i$.

\section{Gaussian basis fusion \label{app:gaussfus}}

The process of creating the Gaussian distionary for the volumetric representation is as follows: each of the atoms in the molecule are assigned a weight and a hydrogen association constant, which is a dimensionless parameter scaling hydrogen's influence on an atom's weight (see Table \ref{tab:gaussification_parameters}) \citep{sigworth_aemcoderepository_2022,kirkland_advanced_1998}.

\begin{table}
    \centering
\begin{tabular}{r r r}
\toprule 
\multicolumn{1}{c}{Element $e$}
& \multicolumn{1}{c}{Atom weight $a_{e}$}
& \multicolumn{1}{c}{Hydrogen Association $b_e$} 
\\
\midrule 
H & 25 & 0.0\\
N & 108 & 1.1\\
C & 130 & 1.3\\
O & 97 & 0.2\\
S & 100 & 0.5\\
P & 267 & 0.0\\
\bottomrule
\end{tabular}
    \caption{Parameters used in creating the Gaussian volumetric representation}
    \label{tab:gaussification_parameters}
\end{table}

At the simplest level, each atom in the molecule can be represented by a Gaussian, with width $1.0$ and weight $a_{e} + b_e a_{H}$ (where $a_{e}$ is atom weight, $b_e$ is hydrogen association, and $a_{H}=25$ is the atom weight of hydrogen). However, this can result in $10^{4}$--$10^{6}$ individual Gaussians. Since cryo-EM images have low SNR and limited resolution, it is possible to simplify this representation by merging the Gaussians. More specifically we use a moment-matching based approach (Algorithm \ref{alg:gauss-merge}) to merge the large number of Gaussians to one per residue.

We approximate each residue by a single Gaussian (the algorithm generalises to higher numbers of Gaussians, but one was found to be adequate), and merging Gaussians with an moment-matching based approximation. This moment-based merging method is used in many Gaussian mixture reduction algorithms (see \citet{crouse_look_2011}).

\begin{algorithm}[H]
\caption{Gaussian merging algorithm}
\label{alg:gauss-merge}
\begin{algorithmic}
    \State \textbf{Input}: A list, for each residue $i$, of atomic positions $\{\mu_{i,j}\}_{i,j}$ and Gaussian amplitudes $\{\alpha_{i,j}\}_{i,j}$, for each atom $j$ in that residue. $N$ the total number of residues, and $N_i$, the number of atoms in the $i$th residue.
    \State Let $\alpha_{i,j}'=\frac{\alpha_{i,j}}{\sum_{i=1}^{N}\sum_{j=1}^{N_i}\alpha_{i,j}}.$
    \For{$i=1\ldots N$}
        \For{$j=1\ldots N_i$}
            \State $\bar{\alpha}_i=\sum_{j=1}^{N_i}\alpha_{i,j}$
            \State $\bar{\mu}_i=\frac{1}{\bar{\alpha}_i}\sum_{j=1}^{N_i}\alpha_{i,j}'\mu_{i,j}$ 
            \State $\bar{\Sigma}_i=\frac{1}{\bar{\alpha}_i}\sum_{j=1}^{N_{i}}\left(\alpha_{i,j}'I+\alpha_{i,j}'(\mu_{i,j}-\bar{\mu}_{i})(\mu_{i,j}-\bar{\mu}_{i})^{T}\right)$
        \EndFor
    \EndFor
    \State \Return{ a list of amplitudes, centres, and width matrices, $\{\bar{\alpha}_i\}_i,\{\bar{\mu}_i\}_i,\{\bar{\Sigma}_i\}_i$}
\end{algorithmic}
\end{algorithm}

To obtain our basis or dictionary $g_i$ for a given protein structure from its protein database~(PDB) data, we use the following process. First the PDB file is parsed to obtain the atomic coordinates, the residues and the elements of each atom. This yields the template backbone $B_0$ (the subscript 0 denotes that it is the known template backbone, calculated from the PDB data, as opposed to the estimated backbone during optimisation). The weights for each atom are then computed using $a_{e} + b_e a_{H}$ and Table \ref{tab:gaussification_parameters}. Using the residue data extracted from the PDB file earlier, we group the atoms by residue, and then apply Algorithm \ref{alg:gauss-merge}. From this we obtain a representation of the volume as a weighted sum of Gaussians 
\[
  \sum_i \bar{\alpha}_i \cdot \exp\left( -\frac{1}{2} (\mathbf{x} - \bar{\mu}_i)^T \bar{\Sigma}_i^{-1} (\mathbf{x} - \bar{\mu}_i) \right),
\]  
where each Gaussian corresponds to a residue. This is the Gaussian volumetric representation of the template backbone $\mathcal{G}(B_0)$.

\subsection{Computation of stored values from $B_0$}

As each of these Gaussians represents the electron density of a residue, the orientation and position of these Gaussians must be defined relative to the backbone conformation. To decouple the position and orientation of the Gaussian from changes in backbone conformation, we use the local frames introduced in \ref{app:relgauss} to express the Gaussian for each residue. This allows us to decouple the parameters of each Gaussian from the conformation. For residue $i$, the local frame is tied to the backbone atoms $\mathbf{r}_{3i}, \mathbf{r}_{3i+1}, \mathbf{r}_{3i+2}$. This local frame has origin at $\mathbf{r}_{3i+1}$ and is defined by the basis vectors $(\mathbf{e}_0, \mathbf{e}_1, \mathbf{e}_2)$, derived from the backbone atoms $\mathbf{r}_{3i}, \mathbf{r}_{3i+1}, \mathbf{r}_{3i+2}$ as described in \ref{app:relgauss}, yielding,
\[
E_{i}=\left[\begin{array}{ccc}
\boldsymbol{e}_{0} & \boldsymbol{e}_{1} & \boldsymbol{e}_{2}\end{array}\right]
\]
The Gaussians can be expressed in the local frame as,
\[
g_{i}^{\text{loc}}(\mathbf{x}')=\bar{\alpha}_{i}\cdot\exp\left(-\frac{1}{2}(\mathbf{x}'-\bar{\mu}_{i}')^{\top}\,(\bar{\Sigma}_{i}')^{-1}\,(\mathbf{x}'-\bar{\mu}_{i}')\right),
\]
where

\begin{align*}
\bar{\mu}_{i}' & =E_{i}(B_0)^{\top}(\bar{\mu}_i - o_{i}(B_0))\\
\bar{\Sigma}_{i}' & =E_{i}(B_0)^{\top}\,\bar{\Sigma}_{i}\,E_{i}(B_0).
\end{align*}

The values of $\{\bar{\alpha}_{i}\},\{\bar{\mu}_{i}'\},\{\bar{\Sigma}_{i}'\}$ are stored in the computer code, and represent the Gaussians in their local frame.

\subsection{Transformation to new conformation $B_\psi$}

When the conformation changes from the template conformation,
the position and orientations of the Gaussians can be computed in the
following way, to match the conformation of the backbone.
Suppose the conformation changes by $\Delta\psi$, making the current
conformation $\psi$. Then the current backbone is $B_{\psi}=\mathcal{W}_{\Delta\psi}(B_{0})$
(the template backbone deformed by $\Delta\psi$). Let:

\[
B_{\psi}=\begin{bmatrix}\bigl(\mathbf{r}_{0}{}^{\psi},\mathbf{r}_{1}{}^{\psi},\mathbf{r}_{2}{}^{\psi}\bigr) & \ldots & \bigl(\mathbf{r}_{3(n-1)}{}^{\psi},\mathbf{r}_{3(n-1)+1}{}^{\psi},\mathbf{r}_{3(n-1)+2}{}^{\psi}\bigr)\end{bmatrix}
\]

From this, we set $E_{i}(B_{\psi})=\left[\begin{array}{ccc}
\boldsymbol{e}_{0}^{\psi} & \boldsymbol{e}_{1}^{\psi} & \boldsymbol{e}_{2}^{\psi}\end{array}\right]$ and $\mathbf{o}_{i}(B_{\psi})=\boldsymbol{r}_{3i+1}^{\psi}$. This is
the local frame for $\psi$ (instead of the template conformation).
The values of $\{\bar{\alpha}_{i}\},\{\bar{\mu}_{i}'\},\{\bar{\Sigma}_{i}'\}$ computed previously stay invariant in their local frames, but need to be computed in the global coordinate system.
To do this we simply transform the parameters of each local Gaussian
dictionary element back to the global frame:

\begin{align*}
  \bar{\mu}_{i}(B_{\psi}) & = \mathbf{o}_{i}(B_{\psi}) + E_{i}(B_{\psi})\,\bar{\mu}_{i}' \\
  \bar{\Sigma}_{i}(B_{\psi}) & = E_{i}(B_{\psi})\,\bar{\Sigma}_{i}'\,(E_{i}(B_{\psi}))^{\top}
\end{align*}

Thus the volumetric representation is the sum of the Gaussian basis
elements given by:
\[
g_{i}(\mathbf{x}, B_{\psi}) = \bar{\alpha}_{i} \cdot \exp\left( 
-\frac{1}{2} 
(
\mathbf{x}-\bar{\mu}_{i}(B_{\psi})
)^{\top} 
(
\bar{\Sigma}_{i}(B_{\psi})
)^{-1} 
(
\mathbf{x}-\bar{\mu}_{i}(B_{\psi})
)
\right).
\]

\section{From volumes to images \label{sec:volimg}}

\begin{figure}[bt]
    \centering
    \includegraphics[width=0.9\linewidth]{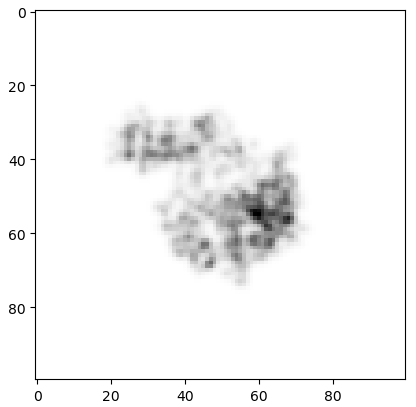}
    \caption{The forward operator is the projection of weighted sum of the Gaussian dictionary functions.}
    \label{fig:fwd-operator}
\end{figure}

Representation in terms of a Gaussian dictionary allows for the simplification and computationally efficient implementation of the forward operator.
In particular, we use the forward operator:
\[ \mathcal{F}_{\mathbf{\omega}}(\mathcal{G}(B))(\mathbf{x}) = 1 - \int^{\infty}_{-\infty} \mathcal{G}(B)(\mathbf{x} + s\mathbf{\omega})\, ds.\]
where $\mathbf{x}\in \mathbb{R}^3$ and $\omega$ is an unit vector denoting the projection direction. This is projection of a weighted sum of Gaussian, which is computationally efficient to calculate. For an example, see Fig.~\ref{fig:fwd-operator}

To summarise all operators defined thus far in this work, we have
\[
  \text{Template Backbone } B_0 \xrightarrow{\mathcal{W}_{\Delta\psi}} B \xrightarrow{\mathcal{G}} \text{Volume} \xrightarrow{\mathcal{F}_{\mathbf{\omega}}} \text{Image}.
\]

Here, we assume that the the prior that is induced by parametrising the backbone chain in terms of rotations and torsions of bonds is strong, and that the effects of noise are far larger than the effects of CTF. A model of the CTF can be easily included if desired, but we have found in our experiments that results have shown robustness to this omission. Future investigation can include investigation of the integration of the CTF (which in most methods is estimated separately)

The representation of the protein chain using Gaussian basis functions resembles the common approach in coarse graining used in molecular dynamic simulations for protein folding \citep{kmiecik_coarse-grained_2016}, where a Gaussian basis function is placed at the corresponding $C_\alpha$ for each residue. This forms a string of ``beads'' where the positions of the protein chain are given by a series of torsion angles. In our version, the Gaussian is not placed at the position of the $C_\alpha$. This is because we are attempting to represent the image through Gaussians, and the side chain of certain residues may be long, placing the the centre of the electron density for those residues at a distance from their $C_\alpha$ atoms. However, the Gaussian positions are still related to the positions of the $C_\alpha$, which are related to other atoms in the backbone chain through torsion angles. 

The approach also bears resemblance to fragmentation methods \citep{gordon_fragmentation_2012}: a class of methods for performing quantum mechanics calculations on larger molecules by explicitly considering only one part (fragment) of the whole in any particular calculation. More precisely, the idea is to divide the molecule of interest into fragments, perform localised quantum calculations on each fragment, and then combine the results from the fragment calculations to predict the same properties for the whole with an accuracy that would be obtained from a full (nonfragmented) calculation. In our case, we divide the protein chain into residues, calculate a Gaussian that represents the electron density for each localised fragment, then combines these into a representation for the whole molecule.

\end{document}